\documentclass[fleqn,usenatbib]{mnras}

\usepackage{array}
\usepackage{framed}
\usepackage{graphicx,epsfig,psfig,multicol}
\usepackage[dvipsnames]{xcolor}
\usepackage[]{inputenc,amssymb}
\usepackage{amsmath}
\usepackage{color}
\usepackage{epsfig}
\usepackage{epstopdf}
\usepackage{textcomp}
\usepackage{xcolor,cancel}
\usepackage{comment}
\usepackage{subcaption}
\usepackage{svg}

\usepackage[export]{adjustbox}
\usepackage{adjustbox}
\usepackage{placeins}

\usepackage{float}
\usepackage{morefloats}

\usepackage{lipsum}
\usepackage{mwe}
\usepackage{arydshln} 

\definecolor{webgreen}{rgb}{0,.5,0}
\definecolor{webbrown}{rgb}{.6,0,0}

\usepackage[toc,page]{appendix}

\def \beq{\begin{equation}}
\def \eeq{\end{equation}}
\def \bea{\begin{eqnarray}}
\def \eea{\end{eqnarray}}

\newcommand{\OmgK}{\mbox{${\Omega}_{\rm K}$}}
\newcommand{\p}{\partial}

\newcommand{\omd}{\mbox{$\omega_{\rm d}$}}
\newcommand{\dt}{\mbox{${\rm d}t$}}

\newcommand{\diff}{\mbox{${\rm d}$}}
\newcommand{\Thphi}{\mbox{$\Theta_\varphi$}}
\newcommand{\Thr}{\mbox{$\Theta_r$}}

\usepackage[T1]{fontenc}

\DeclareRobustCommand{\VAN}[3]{#2}
\let\VANthebibliography\thebibliography
\def\thebibliography{\DeclareRobustCommand{\VAN}[3]{##3}\VANthebibliography}

\title[Resonant Stellar Captures]
{Resonant Capture of Stars by Black Hole Binaries: Extreme Eccentricity Excitation}
\author[O. Reved et al.]{
Omri Reved\thanks{E-mail: \href{mailto:omri.reved@mail.huji.ac.il}{omri.reved@mail.huji.ac.il}}, Lazar Friedland, Nicholas C. Stone
\\
Racah Institute of Physics, The Hebrew University, 91904, Jerusalem, Israel}

\date{Accepted XXX. Received YYY; in original form ZZZ}

\pubyear{2024}

\begin{document}

\label{firstpage}
\pagerange{\pageref{firstpage}--\pageref{lastpage}}
\maketitle

\begin{abstract}
Massive black hole (MBH) binaries in galactic nuclei are one of the leading sources of $\sim$ mHz gravitational waves (GWs) for future missions such as {\it LISA}.  However, the poor sky localization of GW interferometers will make it challenging to identify the host galaxy of MBH mergers absent an electromagnetic counterpart.  One such counterpart is the tidal disruption of a star that has been captured into mean motion resonance with the inspiraling binary.  Here we investigate the production of tidal disruption events (TDEs) through capture into, and subsequent evolution in, orbital resonance.  We examine the full nonlinear evolution of planar autoresonance for stars that lock in to autoresonance with a shrinking MBH binary.  Capture into the 2:1 resonance is guaranteed for any realistic astrophysical parameters (given a relatively small MBH binary mass ratio), and the captured star eventually attains an eccentricity $e\approx 1$, leading to a TDE.  Stellar disks can be produced around MBHs following an active galactic nucleus episode, and we estimate the TDE rates from resonant capture produced when a secondary MBH begins inspiralling through such a disk.  In some cases, the last resonant TDE can occur within a decade of the eventual {\it LISA} signal, helping to localize the GW event.

\end{abstract}



\twocolumn


\section{Introduction}
\label{sec:into}

Astrophysical black holes are divided into three main categories -- stellar mass black holes, intermediate mass black holes (IMBHs), and supermassive black holes (SMBHs). While there is extensive evidence for stellar mass black holes and SMBHs, the existence of IMBHs is still considered controversial, though there is growing observational evidence for them (e.g. \citealt{Greene}).  IMBH candidates are most commonly claimed in dense star systems, either globular clusters or nuclear star clusters, and the search for them is a major challenge in high energy astrophysics, as their mass distribution would encode long-standing questions on the origins of SMBHs \citep{Volonteri10, Inayoshi+20}.

One way to detect massive black holes (MBHs, a catchall term we will use to referr ot both IMBHs and SMBHs), is from electromagnetic flares generated by strong encounters with nearby stars.  Near a MBH in a galactic nucleus, a star can be torn apart by tidal forces, producing a tidal disruption event (TDE). TDEs were first proposed theoretically decades ago \citep{Hills, Rees}, and since then have been observed, both in X-rays \citep{Komossa&Bade,Komossa&Greiner} and in UV and visible light \citep{Gezari1, Gezari2, van_Velzen1}). Their radiation is very bright, characterized by a high-temperature quasi-thermal spectrum, which lasts for months \citep{Arcavi+14, Hung, van_Velzen2} to years \citep{vanVelzen+19}. TDEs are of astrophysical interest for a number of reasons; for example, they can be used as tools to measure black hole mass and spin \citep{Mockler, Wen1, Ryu, Wen2}; they may be the source of the highest-energy neutrinos seen by neutrino astrophysicists \citep{Stein, Reusch}; and lastly, they may be responsible for the growth of IMBHs and SMBHs \citep{Milosavljevic, Stone1}.

A related transient phenomena that occurs in galactic nuclei are EMRIs: extreme mass ratio in-spirals. When a small compact object orbits a massive one (say, a stellar black hole around a SMBH), it emits gravitational waves (GWs), causing its orbit to decay \citep{Ryan,Sigurdsson}. By measuring the GWs emitted during an EMRI, one can perform precision tests of general relativity \citep{Amaro-Seoane1}, although this is ultimately a goal for the future, since the $\sim$mHz GWs produced by EMRIs are too low in frequency to be seen by current ground-based detectors. However, in the near future, space-based laser interferometers such as LISA \citep{Amaro-Seoane2} and Taiji/TianQin \citep{Luo+16, Ruan+20} will observe these GW signals.  EMRI rates are highly uncertain, with current error bars on theoretical rate estimates spanning at least three orders of magnitude \citep{Babak}.  The dynamical processes that produce EMRIs are varied, with multiple possible formation channels \citep{HopmanAlexander05, Miller+05, PanYang21}, but in general EMRIs are much less common than TDEs \citep{Broggi+22}.

In order for either a star to become a TDE or a compact object to become an EMRI, it needs to move into a highly eccentric orbit. In normal galaxies, this process is caused by an ensemble of weak 2-body scatterings from nearby stars, which leads to a random walk in orbital angular momentum \citep{Frank, Cohn}. The rates of both TDEs and EMRIs, in this picture, are set by diffusion through phase space. The characteristic time length for this diffusion is quite long -- in typical galaxies, TDEs are expected once every $\sim 10^4 {\rm yr}$ \citep{Wang, Stone2, van_Velzen3}. 
 Orbital diffusion to produce EMRIs has additional dynamical complexities \citep{BarOrAlexander16, QunbarStone23}, but is still a slow process.  However, when the nucleus of a galaxy undergoes severe dynamical disturbances, the TDE rate can be elevated dramatically; one such example of a disturbance is the formation of a MBH binary.

MBH binaries most typically inspiral and merge in galactic nuclei.  The canonical example of this occurs in the aftermath of a galaxy merger, when two MBHs sink to the center of the merged galaxy under the influence of dynamical friction \citep{Begelman+80}.  Once the MBH binary becomes ``hard,'' i.e. the binary circular speed exceeds the local stellar velocity dispersion, dynamical friction becomes ineffective and the binary stalls for an extended period of time.  As this happens for binary semimajor axes $a\sim {\rm pc}$, this stalling is known as the ``final parsec problem'' \citep{Begelman+80, Merritt&Milosavljevic05}, and is only resolved once external forces reduce $a$ to scales of $\sim 1-10~{\rm mpc}$, where the (circular) GW inspiral time \citep{Peters} is less than a Hubble time.  Possible solutions to the final parsec problem include three-body scatterings with stars \citep{Quinlan96, Milosavljevic&Merritt01}, which achieve high efficiency if the global potential of the galaxy is triaxial \citep{MerrittPoon04, Vasiliev+15}; secular or chaotic interactions with a tertiary MBH brought in by a subsequent galaxy merger \citep{HoffmanLoeb07, Ryu+18, Sayeb+24}; or hydrodynamic torques from a circumbinary accretion disk \citep{GouldRix00, Hayasaki+07, Cuadra+09} during an episode of active galactic nucleus (AGN) activity.  

All of these solutions to the final parsec problem will ultimately produce a compact MBH binary, where each MBH is orbited by a remnant of its original bound cloud of stars.  As the MBH binary inspirals under gravitational radiation reaction, many of these stars will be captured into mean motion resonance, and may evolve to higher eccentricity orbits as we have studied.  However, this scenario is not immediately describable with the formalism we will develop below, as the nuclear star clusters surrounding MBHs are quasi-spherical \citep{Neumayer+20}, adding extra degrees of freedom to the 2D, planar autoresonance dynamics that we will quantify\footnote{Of course, a very small number of stars from the quasi-spherical nuclear star cluster will be chance lie in the orbital plane of the MBH binary, and their orbital evolution can be described by our formalism.}.  A more promising way for classic MBH binaries to produce a large population of coplanar stars is in the framework of hydrodynamic solutions to the final parsec problem.  In order for a circumbinary gas disk to harden an MBH binary to GW scales, the disk itself must be so massive as to be self-gravitating, which will likely lead to substantial star formation \citep{Lodato+09}.  

Self-gravitating disks are generally Toomre unstable \citep{Toomre}, meaning that small density fluctuations will run away and produce compact, gravitationally bound fragments.  In the context of AGN gas disks, analytic arguments \citep{Sirko, Thompson}, numerical hydrodynamic simulations \citep{Nayakshin+07}, and observations of our own Galactic Center \citep{Levin&Beloborodov, Paumard+06} all strongly indicate that the outcome of this process is the formation of a disk of stars, possibly with a top-heavy initial mass function \citep{Bartko+10, Lu+13}.  Star formation in the circumbinary disk will produce a population of exterior stars that cannot be captured into autoresonance, but star formation in circumprimary disks (fed by the circumbinary disk) will produce planar stellar disks whose components can be captured into resonance as the MBH binary shrinks, as in our scenario above.

A second way to produce an MBH binary in a galactic nucleus begins with a Toomre unstable AGN disk around a {\it single} MBH.  Individual stars in such an environment may grow to a supermassive size, terminating their lives as IMBHs \citep{GoodmanTan04} that can undergo subsequent migration inwards towards the central MBH.  IMBHs may also be assembled through repeated mergers of stellar mass black holes in AGN disks \citep{McKernan+11, Secunda+20}.  Other stars that exist at smaller radii in such a disk may then be captured into resonance.  Unlike in the first scenario, this one is unlikely to feature stellar disks around both MBHs, but only around the primary.  In both scenarios, the vacuum dynamics of autoresonance we have studied here will only apply once the AGN gas has dissipated.

Afterwards, we are left with a disc of stars, and perhaps an IMBH. If an IMBH exists in one of these stellar disk, it can in-spiral inwards, due to gravitational wave emission or dynamical friction \citep{Amaro-Seoane1}. This in-spiralling may cause capture into resonance of nearby stars, due to the mechanism of autoresonance -- similar to what happened in to Plutinos in the early Solar system, due to Neptune's changing orbit \citep{Friedland, Yu}. Autoresonance has been proposed to bring some stars closer to SMBH-IMBH pairs during their in-spiral \citep{Seto&Muto1, Seto&Muto2}, but the problem has not previously been investigated systematically\footnote{The related phenomenon of resonance capture in the more complicated early (i.e. gas-rich) stages of an AGN disk has recently been examined numerically by \citet{Secunda+19, PengChen23}.}.

In this paper, we will study the problem of capturing into mean motion resonance a star by two co-orbiting black holes: a supermassive, and an intermediate mass one. In Sec. 2, we will begin by examining the phenomena qualitatively, by looking at orbital simulations. In Sec. 3, we will develop an analytical theory for the capture into resonance of the system, and validate our results by comparing them to numerical simulation. In Sec. 4, we will study the problem at the developed stage, after the capture into resonance. Lastly, in Sec. 5, we will look at the astrophysical consequences of the theory.

\section{Autoresonance in Orbital Simulations}

\label{sec:numerics}

\begin{figure}
\centering
\includegraphics[width=0.3\textwidth]{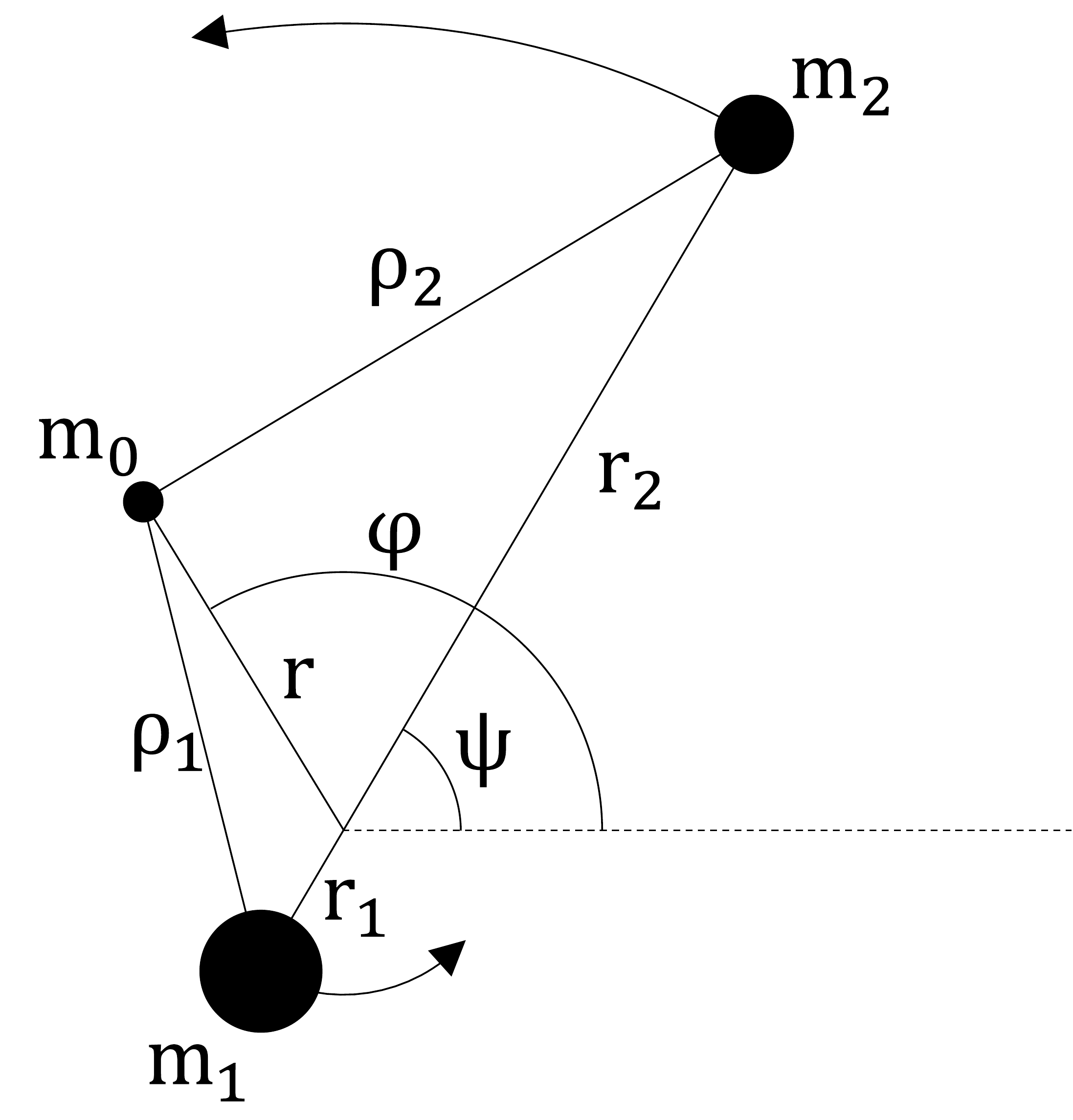}
\caption{Geometry of the adiabatic-restricted three-body problem. The
primary and secondary masses $m_1$ and $m_2$ move on contracting, spiralling
orbits, having a slowly varying local angular frequency $\protect\omega_d(t)=
\mbox{${\rm d}$}\protect\psi/\mbox{${\rm d}t$}$.}
\label{fig:sys_description}
\end{figure}

Consider the planar hierarchical restricted three-body problem: a test mass $%
m_{0}$ (a star) affected by two dominant masses $m_{1,2}\gg m_{0}$ (a BH
binary), co-rotating unperturbed by the test mass on circular orbits around
their center of mass,  as illustrated in figure \ref{fig:sys_description}.
Assume that the dominant masses slowly loose their energy, and therefore
spiral inwards. The Hamiltonian of this system can be written as
\begin{equation}
{}H=\frac{1}{2}\left( p_{r}^{2}+\frac{p_{\varphi }^{2}}{r^{2}}\right) -\frac{%
1}{\rho _{1}}-\frac{q}{\rho _{2}}, \label{full_Hamiltonian}
\end{equation}%
where $\rho _{1}^{2}=r^{2}+r_{1}^{2}+2rr_{1}\cos{(\varphi -\psi )}$ and $%
\rho _{2}^{2}=r^{2}+r_{2}^{2}-2rr_{2}\cos{(\varphi -\psi )}$ are the distances
from the dominant masses,   $m_{1}r_{1}=m_{2}r_{2}$, $\psi (t)$ is the
rotation angle of $m_{1,2}$, and $q=m_{2}/m_{1}$. We will consider the case
in which the test mass starts on a circular orbit of radius $r_{0}$ at the
angular frequency $\omega _{0}\approx (Gm_{1}/r_{0}^{3})^{1/2}$. Note that
in equation \eqref{full_Hamiltonian}, $Gm_{1}$ is replaced by unity. This can
be done by using dimensionless time $t\rightarrow \omega _{0}t$,
dimensionless distances $r\rightarrow r/r_{0}$ and $\rho
_{1,2}\rightarrow \rho _{1,2}/r_{0}$, and dimensionless momenta: $p_{\varphi
}\rightarrow p_{\varphi }/p_{\varphi, 0}$ and $p_{r}\rightarrow
p_{r}r_{0}/p_{\varphi, 0}$, where $p_{\varphi, 0}=\omega _{0}^{2}r_{0}$.
As mentioned above, we will assume that $\omd=\mbox{${\rm
d}$}\psi /\mbox{${\rm d}t$}$ is a slow function of time, i.e. $%
\mbox{$\omega_{\rm d}$}^{-2}A\ll 1$, where $A=\mbox{${\rm d}$}%
\mbox{$\omega_{\rm d}$}/\mbox{${\rm d}t$}>0$. We will also assume that the
mass ratio $q$ is a small parameter ($q\ll 1$), as some aspects of our theoretical modeling will break down as $q\to 1$.  At the most basic level, resonance overlap begins to severely destabilize individual mean motion resonances for large $q$ values.  Both analytic resonance overlap criteria \citep{Wisdom80} and numerical orbit integrations \citep{HolmanWiegert99, MudrykWu06} show that when $q \gtrsim 0.01$, the 2:1 mean motion resonance becomes unstable, and stars that are temporarily placed in it will quickly escape due to chaotic orbital evolution.

\begin{figure}
\centering
\includegraphics[width=0.5\textwidth]{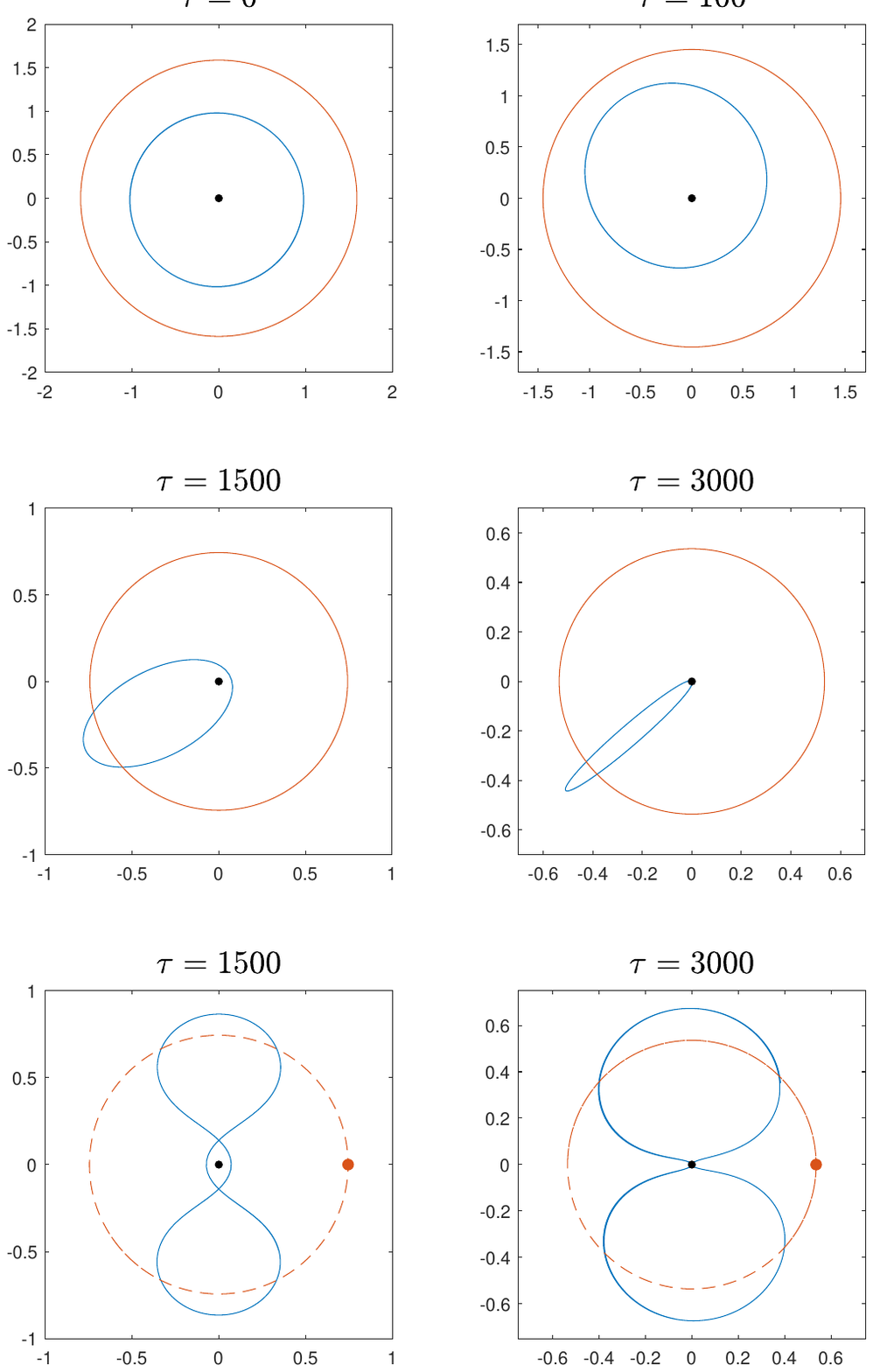}
\caption{A typical solution of the equations of motion derived from the
Hamiltonian at equation \eqref{full_Hamiltonian}, for mass ratio $q=10^{-4}$%
, with linear chirp rate of $10^{-6}$. The orbit of $m_2$ is portrayed in
orange, while the test mass' one is in blue. The two lower panels show the
situation from a co-rotating frame, where the secondary black hole $m_2$
located on the right.}
\label{fig:frames}
\end{figure}

\begin{figure}
\centering
\includegraphics[width=0.5%
\textwidth]{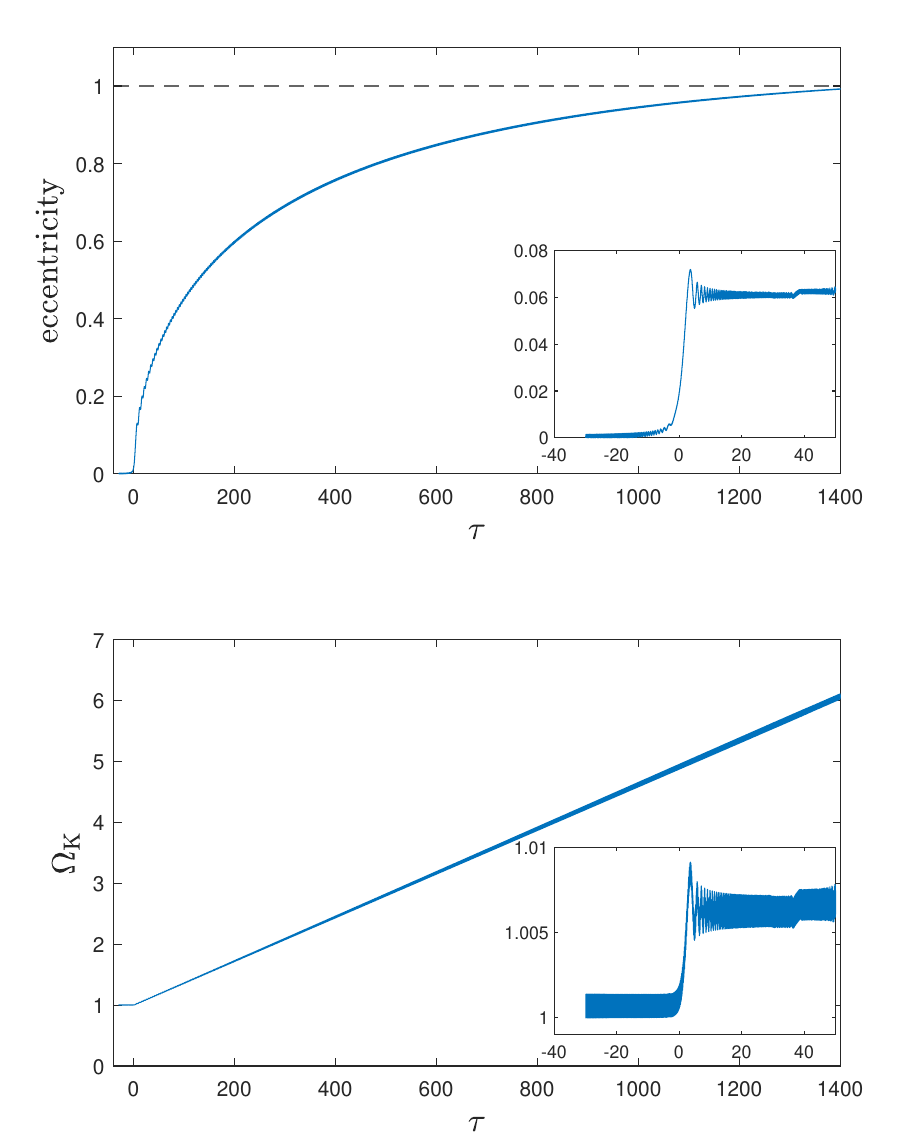}
\caption{A typical solution of the equations of motion derived from the
Hamiltonian at equation \eqref{full_Hamiltonian}, for mass ratio $q=10^{-4}$%
; on the upper panel, the eccentricity of test mass' orbit is shown, while
on the lower one the angular frequency of the test mass, $
\mbox{${\Omega}_{\rm K}$}$ is plotted. On the large panels, the solution for
linear chirp rate of $A=6.5\cdot10^{-6}$, while the smaller ones shows the
solution for $A=8\cdot10^{-6}$.}
\label{fig:eccentricity_freq}
\end{figure}

We illustrate the process of resonant capture of the test mass into $2:1$ resonance
and continuing autoresonance in this system in numerical examples in figures \ref{fig:frames} and \ref{fig:eccentricity_freq} for the case of $q=10^{-4}$, chirped rotation frequency $%
\omega _{d}={\rm d}\psi /\dt=1/2+At$ and constant chirp rate $A=10^{-6}$. The
resonance passage in this example takes place at $t=0$ and four upper panels
in Fig. 2 show the orbits of the three masses at different times (the slow time $\tau =\sqrt{2A}t$ is used in the figures). One observes that
the eccentricity of the test mass increases continuously reaching near unity
at $\tau =3000$. The lower two panels in Fig. 2 show the trajectories in the
co-rotating plane and illustrate that the test mass closely approaches, but
never collides with the secondary mass $m_{2}$.  Additional
simulation results are presented in Fig. 3 for larger $A=6.5\times 10^{-6}$
showing the eccentricity of the test mass and its Keplerian $\Omega _{K}$
frequency as functions of the slow time  $\tau $. One can see that again the
eccentricity of the mass approaches unity, while the Keplerian frequency
closely follows $2\omega _{d}$ in time constituting the continuing $2:1$
autoresonance in the system. Nevertheless, for even larger chirp rate, $%
A=8\times 10^{-6}$ (inner panels in Fig. 3) the resonance is destroyed close to
$\tau =0$ and the eccentricity remains small. We find that this loss of
autoresonance in our system takes place if $A$ exceeds some threshold
value scaling as $A_{\rm th}\sim q^{4/3}$and discuss this phenomenon next.

\section{Capture into Autoresonance and the Threshold Phenomenon}

\label{sec:capture}

In analysing our system, we expand $1/\rho _{1}$ and $q/\rho _{2}$ in equation %
\eqref{full_Hamiltonian} to first order in $q$ and rewrite $H=H_{0}(r,p_{r},p_{\varphi })+qV_{1}(r,\theta )$, where%
\begin{equation}
H_{0}=\frac{1}{2}\left( p_{r}^{2}+\frac{p_{\varphi }^{2}}{r^{2}}\right) -%
\frac{1}{r}  \label{H0}
\end{equation}%
is the unperturbed Hamiltonian and
\begin{equation}
r_{2}V_{1}=\alpha ^{2}\cos {\theta }-\frac{\alpha }{\sqrt{1+\alpha
^{2}-2\alpha \cos {\theta }}},  \label{per_poten}
\end{equation}%
with $\alpha =r_{2}/r$ and $\theta =\varphi -\psi $. The following analysis
is similar to that in \cite{Friedland} for the problem of formation of
Plutinos in the early stages of evolution of the solar system, but in
contrast to Plutinos we consider the case of inner resonances, i.e. $\alpha
(t=0)>1$.

We can rewrite equation \eqref{per_poten} as an expansion in harmonics of $\theta
$:
\begin{equation}
r_{2}V_{1}=\sum_{j=0}^{\infty }f_{j}(\alpha )\cos {\left(j\theta\right)},
\label{per_poten_harm}
\end{equation}%
where $f_{i}$ are expressed in terms of the Laplace's coefficients $%
b_{1/2}^{(j)}(\alpha )$ \citep{Tremaine}:
\begin{equation}
{}f_{j}(\alpha )=%
\begin{cases}
-\frac{\alpha }{2}b_{1/2}^{(0)}(\alpha ) & j=0 \\
\alpha ^{2}-\alpha b_{1/2}^{(1)}(\alpha ) & j=1 \\
-\alpha b_{1/2}^{(j)}(\alpha ) & j\geq 2%
\end{cases}%
.  \label{fj}
\end{equation}%

The next step is to transform the problem to the action-angle (AA) variables of the unperturbed Hamiltonian (see Appendix A
for details): $(p_{r},p_{\varphi },r,\varphi)\rightarrow (J_{r},J_{\varphi },\Theta _{r},\Theta _{\varphi })$, and to use single resonance approximation \citep{Chirikov} (see Appendix C). This yields the approximate resonant Hamiltonian for the $p:p-1$ resonance of the form
\begin{equation}
{}H=H_{0}(J_{r},J_{\varphi })+qV_{p}(J_{r},\Phi ),
\label{perturbed_hamiltonian}
\end{equation}%
where
\begin{equation}
H_{0}=-\frac{1}{2(J_{r}+J_{\varphi })^{2}}  \label{H0JrJphi}
\end{equation}
and $\Phi =p\mbox{$\Theta_\varphi$}-\mbox{$\Theta_r$}-p\psi -\pi $ is the
resonant angle mismatch ($\pi $ was subtracted for convenience as explained
below). The unperturbed Hamiltonian $H_{0}$ yields the Kepler frequency of
the unperturbed test mass%
\begin{equation}
\Omega _{K}=\frac{\diff\Theta _{r}}{\dt}=\frac{\diff\Theta _{\varphi }}{\dt}=\frac{1}{%
(J_{r}+J_{\varphi })^{3}}  \label{Omega_K_AAV}
\end{equation}%
Note that the vicinity of $p:p-1$ resonance means that
\begin{equation}
{}(p-1)\mbox{${\Omega}_{\rm K}$}-p\mbox{$\omega_{\rm d}$}\approx 0,
\label{res_cond1}
\end{equation}%
i.e., in resonance, the mismatch angle $\Phi $ is a slow variable.

In this section we will focus on the initial stage of the resonant capture, i.e.
the case of small eccentricities (small $J_{r})$. As shown in Appendix \ref{app:small_AR} for this case, the resonant perturbing potential $V_{p}$ assumes the following approximate form
\begin{equation}
{}V_{p}\approx -\zeta _{p}\sqrt{2J_{r}}\cos {\Phi },  \label{Vpsmall}
\end{equation}%
where for the first three resonances $\zeta _{2}=0.750$, $\zeta _{3}=1.546
$, and $\zeta _{4}=2.345$. Then, equation \eqref{perturbed_hamiltonian} yields the following
evolution equations
\begin{eqnarray}  
\frac{\mbox{${\rm d}$}J_{r}}{\mbox{${\rm d}t$}} &=&-q\zeta _{p}\sqrt{2J_{r}%
}\sin {\Phi }  \label{dJrdt} \\
\frac{\mbox{${\rm d}$}J_{\varphi }}{\mbox{${\rm d}t$}} &=&pq\zeta _{p}\sqrt{%
2J_{r}}\sin {\Phi }  \label{dJphidt} \\
\frac{\mbox{${\rm d}$}\Phi }{\mbox{${\rm d}t$}} &=&(p-1)\mbox{${\Omega}_{\rm
K}$}-\frac{q\zeta _{p}}{\sqrt{2J_{r}}}\cos {\Phi }-p\omd.
\label{dPhidt1}
\end{eqnarray}%
The first two of these equations give the conservation law (using the initial values $J_{r}=0$, and $J_{\varphi }=1$)
\begin{equation}
pJ_{r}+J_{\varphi }=1.  \label{app_cons}
\end{equation}%
Therefore, the whole system can be described by $J_{r}$ and $\Phi $ only.

We will focus on passage through $p-1:p$ resonance at $t=0$, i.e., write
$\mbox{$\omega_{\rm d}$}\approx (p-1)/p+At$,
where $A$ is the local driving frequency chirp rate at the. Furthermore, assuming small $J_{r}$, one can use Eqs. \eqref{Omega_K_AAV} and %
\eqref{app_cons} to approximate $\mbox{${\Omega}_{\rm K}$}\approx
1+3(p-1)J_{r}$. Then equation \eqref{dPhidt1} becomes \
\begin{equation}
\frac{\mbox{${\rm d}$}\Phi }{\mbox{${\rm d}t$}}=-pAt+3(p-1)^{2}J_{r}-\frac{%
q\zeta _{p}}{\sqrt{2J_{r}}}\cos {\Phi },  \label{EomPhi_approx}
\end{equation}%
and at negative times prior passage through resonance when $%
J_{r}$ is sufficiently small,%
\begin{equation}
\frac{\mbox{${\rm d}$}\Phi }{\mbox{${\rm d}t$}}\approx -pAt-\frac{q\zeta _{p}%
}{\sqrt{2J_{r}}}\cos {\Phi }.  \label{pror}
\end{equation}%
It is the singularity at $J_{r}\rightarrow 0$ in the last term in this
equation which yields efficient phase-locking in the system near the fixed
point ${\Phi =0}$, $\sqrt{2J_{r}}=-\frac{pAt}{q\zeta _{p}}$ at negative
times. After the passage through resonance, $J_{r}$
automatically adjusts in time to stay in the continuing resonance \cite{Friedland}, while $\Phi $ remains small and oscillates around 0 (this was the
reason for subtracting $\pi $ in the definition of $\Phi $ above).

\begin{figure}
\centering
\includegraphics[width=0.5\textwidth]{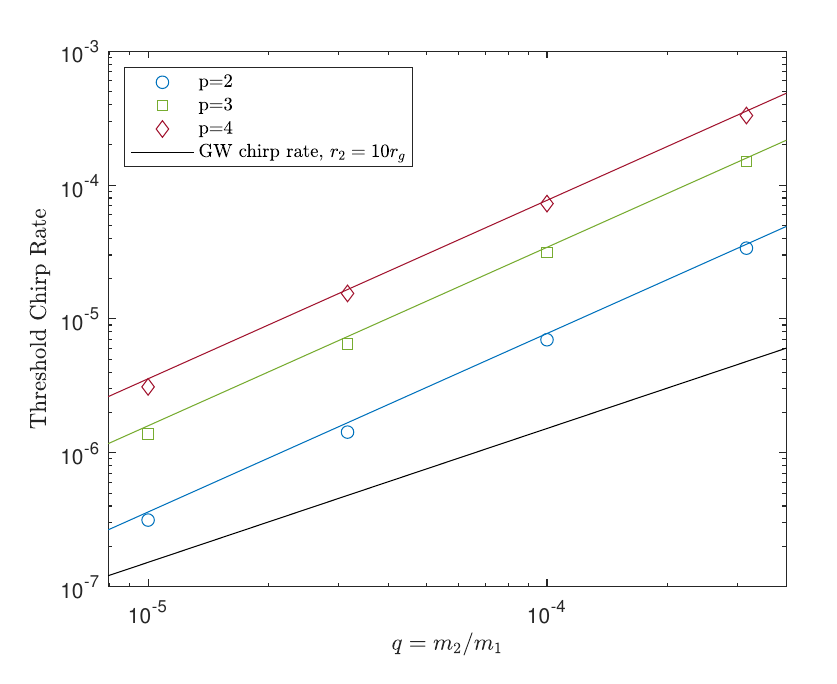}
\caption{A plot showing different thresholds found numerically (markers) and
the theoretical predictions (solid lines). The plots shows the thresholds
for different $p:p-1$ resonances, as dependent by the mass ratio $q$. The
black line demonstrates the gravitational waves chirp rate (dimensionless),
as can be found by \eqref{A_GW}, for $r_2=10r_g$.}
\label{fig:thresholds}
\end{figure}

Equations \eqref{dJrdt} and \eqref{EomPhi_approx} comprise a standard
set describing the autoresonance phenomenon \citep{Friedland_web},  which can
be conveniently simplified by defining $\Delta
=(3J_{r})^{1/2}(pA)^{-1/4}(p-1)$, the slow time $\tau =\sqrt{pA}t$%
, and $\varepsilon =(1.5)^{1/2}q\zeta _{p}(pA)^{-3/4}(p-1)$. With these
definitions, Eqs. \eqref{dJrdt} and \eqref{EomPhi_approx} yield a
single parameter system 
\begin{eqnarray}  \label{approx_Phi_EoM}
\frac{\mbox{${\rm d}$}\Delta }{\mbox{${\rm d}$}\tau } &=&-\varepsilon \sin
{\Phi }  \label{approx_Delta_EoM} \\
\frac{\mbox{${\rm d}$}\Phi }{\mbox{${\rm d}$}\tau } &=&\Delta ^{2}-\tau -%
\frac{\varepsilon }{\Delta }\cos {\Phi }.  \label{Phi_EoM}
\end{eqnarray}%
This system can be rewritten as a single complex nonlinear
Schrodinger-type equation by defining $\Psi =\Delta \exp {(i\Phi )}$:
\begin{equation}
{}i\frac{d\Psi }{dt}+\Psi (|\Psi |^{2}-\tau )=\varepsilon .
\label{standard NLS}
\end{equation}%
One finds that by starting with $\Psi=0$ at negative times \citep{Friedland_web}, this equation yields a sharp threshold phenomena: above a critical
value $\varepsilon _{\rm th}\approx 0.41
$, the system is excited to large amplitudes $\Delta $ at positive times, while ${\Phi }$
remains small. In contrast, below this threshold, $\Delta$ remains small while ${\Phi }$
diverges. Therefore, the test mass in our problem is captured into autoresonance
if $\varepsilon >\varepsilon _{th}$, or
\begin{equation}
A<A_{\rm th}=\frac{3^{2/3}(p-1)^{4/3}\zeta _{p}^{4/3}}{2^{2/3}p\varepsilon_{\rm th}^{4/3}}%
q^{4/3}=\sigma _{p}q^{4/3},  \label{A_th}
\end{equation}%
where $\sigma _{2}=1.567$, $\sigma _{3}=6.902$, and $\sigma _{4}=15.485$. 
The transition from above to below the threshold is seen in the example in figure 
\ref{fig:eccentricity_freq}. One also observes a good agreement between the theoretical and numerical thresholds
$A_{\rm th}$ for different $p$ and $q$ as illustrated in figure \ref{fig:thresholds}. 

Since we consider chirp rates caused by gravitational waves, we have \citep{Peters}
\beq{}
    \label{A_GW}
    A_{\rm GW} = \frac{96}{5}\frac{G^{7/2}m_1m_2(m_1+m_2)^{3/2}}{c^5 r_2^{11/2}}\left(\frac{Gm_1}{r^3}\right)^{-1},\eeq{}where the factor $\Omega_{K}^{-2}$ was added to make $A_{\rm GW}$ dimensionless. By equating Eqs. \eqref{A_th} and \eqref{A_GW}, one gets the condition for capture into resonance 
\beq{}
    \frac{r_2}{r_g}>\left[\frac{96}{5\sigma_pq^{1/3}}\left(\frac{r}{r_g}\right)^3\right]^{2/11},
\eeq{}where $r_g=Gm_1/c^2$ is the gravitational radius of $m_1$. When the capture into resonance occurs, $(p-1)\OmgK\approx p\omd$. Therefore, after some algebra one gets
\beq{}
    \frac{r_2}{r_g}>3.26\left(\frac{\sigma_p}{10}\right)^{-2/5}\left(\frac{q}{10^{-3}}\right)^{-2/15}\left(\frac{p-1}{p}\right)^{4/5},
\eeq{}
which means that the capture into resonance is almost always guaranteed, as the right hand side of this inequality is always $\sim \mathcal{O}(1)$, and realistic IMBH-SMBH inspirals begin with $r_2 \gg r_{\rm g}$. 

Note that as  the phase locking (autoresonance) in the system continues, the resulting growth of $J_r$ and the decrease of $J_{\varphi}=1-pJ_r$ yield a continuing growth of the orbital eccentricity 
\begin{equation}
e=\sqrt{1-\left( \frac{J_{\varphi }}{%
J_{\varphi }+J_{r}}\right) ^{2}.}
\label{ecc}
\end{equation}
Most importantly, this phase-locking and the growth of the eccentricity towards $e=1$  continue beyond the small $J_r$ regime, as will be shown in the next section. 

\section{Developed Autoresonance Stage and Stability}

\begin{figure}
\centering
\includegraphics[width=0.5\textwidth]{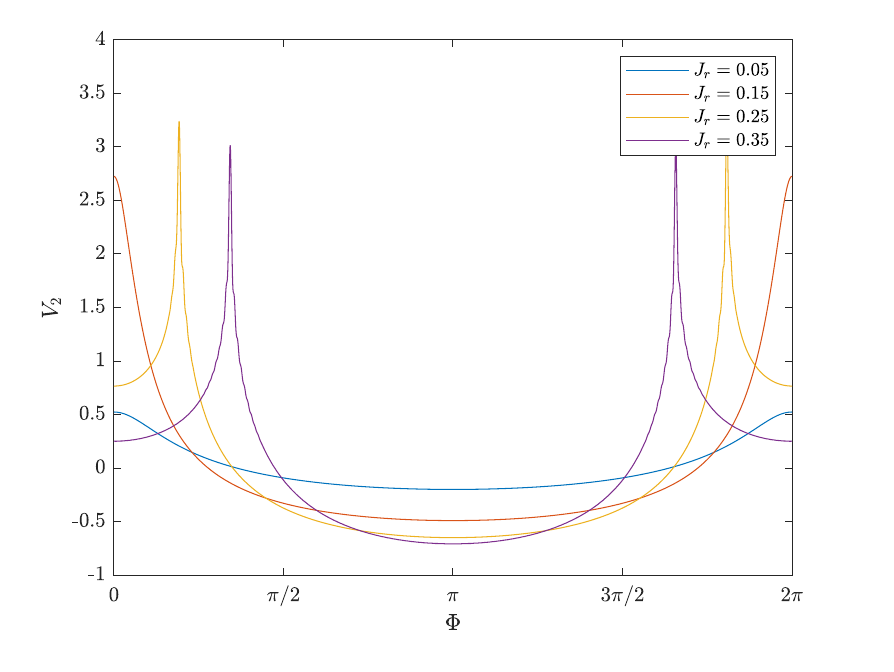}
\caption{The perturbing potential $V_p(J_r,\Phi)$ for $p=2$, at various $J_r$
values.}
\label{fig:potential}
\end{figure}

The analysis of the developed autoresonance stage proceeds from the extension of
the resonant potential $V_{p}$ in equation \eqref{perturbed_hamiltonian} from equation
\eqref{Vpsmall} for small $J_{r}$ to larger values of $J_{r}$ (larger
eccentricities). This generalization
is developed in Appendix \ref{app:potential_derivation} using action-angle variables. Unfortunately, $V_p$ doesn't have a simple analytic form for large $J_{r}$, but can be found numerically. Figure %
\ref{fig:potential} shows this perturbing potential for $p=2$ and several
values of $J_{r}$. The potential still has its minimum at $\Phi =0$, but a
more complex form. In particular, one observes two sharp spikes, starting at
$J_{r}\approx 0.162$. We find that these peaks appear as the orbits of the test mass $m_{0}$ and the secondary BH $m_{2}$ cross. The value of $J_{r}$ for the first such crossing can be found by equating the radius of
the outer black hole in autoresonance
\begin{equation*}
r_{2}=\omega _{\rm d}^{-2/3}=\left( \frac{p-1}{p}\Omega _{K}\right)
^{-2/3}=w(J_{\varphi }+J_{r})^{2}
\end{equation*}%
where $w=\left[ p/(p-1)\right] ^{2/3}$, to the apoapsis distance of
the test mass, given by
\begin{equation}
{}r_{\rm apo}=\frac{J_{\varphi }^{2}}{1-e},
\end{equation}%
where $e$ is the eccentricity (see Eq. \eqref{ecc}). This comparison and use of the
conservation law \eqref{app_cons}  (the same as for small eccentricities,
see below) yields $J_{r}$ at the crossing of the trajectories
\begin{equation}
{}J_{r}^{\rm cr}=\frac{\sqrt{w(2-w)}-1}{(p-1)\sqrt{w(2-w)}-p}.
\end{equation}%
{}

Now, similar to the small $J_{r}$ case, we write the equations of motion:
\begin{eqnarray}
{}\frac{\mbox{${\rm d}$}J_{r}}{\mbox{${\rm d}t$}} &=&-q\frac{\partial V_{p}}{%
\partial \Phi }  \label{EoMJr} \\
\frac{\mbox{${\rm d}$}J_{\varphi }}{\mbox{${\rm d}t$}} &=&pq\frac{\partial
V_{p}}{\partial \Phi } \\
\frac{\mbox{${\rm d}$}\Phi }{\mbox{${\rm d}t$}} &=&(p-1)\mbox{${\Omega}_{\rm
K}$}-\frac{\partial V_{p}}{\partial J_{r}}-p\omega _{d}   \label{EoMPhi}
\end{eqnarray}%
yielding the same conservation law
\begin{equation}
{}pJ_{r}+J_{\varphi }=1.  \label{cons}
\end{equation}%
{} Assuming that following the initial phase-locking stage, the system stays
at small values of mismatch $\Phi $, we expand the perturbing potential $V_{p}
$ to second order in $\Phi $, i.e., write
\begin{equation}
{}V_{p}=U_{0}(J_{r})+\frac{1}{2}k(J_{r})\Phi ^{2},  \label{poten_2nd}
\end{equation}%
which after substitution into Eqs. \eqref{EoMJr} and \eqref{EoMPhi}, yields
\begin{eqnarray}
{}\frac{\mbox{${\rm d}$}J_{r}}{\mbox{${\rm d}t$}} &=&-qk\Phi ^{{}}
\label{dJdt} \\
\frac{\mbox{${\rm d}$}\Phi }{\mbox{${\rm d}t$}} &=&(p-1)\mbox{${\Omega}_{\rm
K}$}-p\omega _{\rm d}-q\left( \frac{\partial U_{0}}{\partial Jr}+\frac{1}{2}   \label{dPhidt} %
\frac{\partial k}{\partial Jr}\Phi ^{2}\right) .
\end{eqnarray}%
{} At this stage, we neglect the last $\mathcal{O}(q)$ term in equation \eqref{dPhidt}.
The reason for this approximation will be explained at the end of this
section. We seek solutions of the resulting system in the form $%
J_{r}=J_{r0}+\delta J_{r}$ and $\Phi =\Phi _{0}+\delta \Phi $ where $\delta
J_{r}$ and $\delta \Phi $ are small and \textit{rapidly} oscillating
components, while $J_{r0}$ and $\Phi _{0}$ are \textit{slow} monotonic
averages, which do not vary significantly during a single oscillation of $%
\delta J_{r}$ and $\delta \Phi $. By separating the oscillating and
monotonic parts in Eqs. \eqref{dJdt}, \eqref{dPhidt}, we get

\begin{eqnarray}
\frac{\mbox{${\rm d}$}\delta J_{r}}{\mbox{${\rm d}t$}} &=&-qk(J_{r0})\delta
\Phi ,  \label{A} \\
\frac{\mbox{${\rm d}$}\delta \Phi }{\mbox{${\rm d}t$}} &=&(p-1)\frac{%
\partial \Omega _{K}}{\partial J_{r0}}\delta J_{r}  \label{B}
\end{eqnarray}%
and%
\begin{eqnarray}
\frac{\mbox{${\rm d}$}J_{r0}}{\mbox{${\rm d}t$}} &=&-qk(J_{r0})\Phi _{0},
\label{C} \\
\frac{\mbox{${\rm d}$}\Phi _{0}}{\mbox{${\rm d}t$}} &=&(p-1)%
\mbox{${\Omega}_{\rm K}$}(J_{r0})-p\omega _{\rm d}.  \label{D}
\end{eqnarray}%
The analysis of these systems proceeds from Eqs. \eqref{C} and \eqref{D}%
. We differentiate equation \eqref{D} and substitute equation \eqref{C} to
get%
\begin{equation}
\frac{d^{2}\Phi _{0}}{dt^{2}}=-q(p-1)k(J_{r0})\frac{\partial %
\mbox{${\Omega}_{\rm K}$}}{\partial J_{r0}}\Phi _{0}-pA.  \label{d2Phidt2}
\end{equation}%
The approximate solution of this equation is obtained by neglecting the LHS,
yielding%
\begin{equation}
\Phi _{0}=-\frac{pA}{q(p-1)k(J_{r0})\frac{\partial \Omega _{K}}{\partial
J_{r0}}}  \label{Phi0}
\end{equation}%
The smallness and slowness of this solution is guaranteed if $A/q\ll 1$
which is assumed in the following. Now we can also neglect the LHS in %
\eqref{D} to write%
\begin{equation}
(p-1)\Omega _{K}(J_{r0})-p\omega _{\rm d}\approx 0,  \label{resonance}
\end{equation}%
which constitutes the continuing resonance (autoresonance) condition.
Finally, we proceed to the oscillating components $\delta J_{r}$ and $\delta
\Phi$ of the solution described by \eqref{A} and \eqref{B}. We observe that
this is a linear oscillatory Hamiltonian system governed by Hamiltonian%
\begin{equation}
h(\delta J_{r},\delta \Phi ^{\prime },t)=\frac{1}{2}\left[ qk(J_{r0})\delta
\Phi ^{2}+(p-1)\frac{\partial \Omega _{K}}{\partial J_{r0}}\delta J_{r}^{2}%
\right] ,  \label{h}
\end{equation}%
where the \textit{slow} time dependence enters via coefficients involving $%
J_{r0}(t)$. For a frozen slow time the trajectory of this system in phase
space is an ellipse%
\begin{equation}
\frac{\delta \Phi ^{2}}{\mathbf{A}^{2}}+\frac{\delta J_{r}^{2}}{\mathbf{B}%
^{2}}=1,  \label{ellipse}
\end{equation}%
where
\begin{equation}
\mathbf{A}=\sqrt{\frac{2h}{qk(J_{r0})}},\mathbf{B}=\sqrt{\frac{2h}{(p-1)%
\frac{\partial \Omega _{K}}{\partial J_{r0}}}}.  \label{axis}
\end{equation}%
Because of the slow time dependence of the parameters $\mathbf{A}$ and $%
\mathbf{B}$ we can use the adiabatic theory \citep{Landau&Lifshitz} yielding the adiabatic invariant  -- the action -- $I=S/(2\pi )$, where $S=\pi \mathbf{AB}$
is the area of the ellipse. Thus,%
\begin{equation}
I=\frac{h}{\Omega _{A}}  \label{Action}
\end{equation}%
where
\begin{equation}
\Omega _{A}=\sqrt{(p-1)qk(J_{r0})\frac{\partial \Omega _{K}}{\partial J_{r0}}%
}  \label{autoresonant frequency}
\end{equation}%
is the frequency of the (autoresonant) oscillations. One can write $%
h=I\Omega _{A}=qk(J_{r0})a^{2}/2=(p-1)\left(\partial \Omega _{K}/\partial J_{r0}\right)\left(b^{2}/2\right)$, where $a$ and $b$ are the amplitudes of oscillations of $\delta
\Phi $ and $\delta J_{r}$ respectively. Therefore, given the adiabatic
invariant $I$ (by initial conditions), we find the time evolution of the
amplitudes:%
\begin{equation}
a^{2}=2I\sqrt{\frac{(p-1)\frac{\partial \Omega _{K}}{\partial J_{r0}}}{%
qk(J_{r0})}}  \label{amplitude}
\end{equation}%
and $b=2I/a$. Note that both $a$ and $b$ remain small for sufficiently small
$q$, meaning that the stability of the autoresonant system is guaranteed. Finally, we write the adiabaticity
condition in our problem 
\begin{equation}
\frac{1}{\Omega _{A}^{2}}\frac{\mbox{${\rm d}$}\Omega _{A}}{\mbox{${\rm d}t$}%
}\ll 1  \label{Adiabaticity condition}.
\end{equation}%
Here equation \eqref{resonance} yields \ $\frac{\mbox{${\rm d}$}\Omega_{A}}{%
\mbox{${\rm d}t$}}\sim \frac{\mbox{${\rm d}$}J_{r0}}{\mbox{${\rm d}t$}}\sim
\mathcal{O}(A)$, while $\Omega _{A}^{2}\sim q$. Therefore, condition  $A/q\ll 1$ guaranties the adiabaticity. Note that this is the same
condition used above when studying the evolution of the slow system \eqref{C}
and \eqref{D}. Finally, note that the RHS in equation \eqref{dPhidt}
evolves as $(p-1)\frac{\partial \Omega _{K}}{\partial J_{r0}}\delta
J_{r}\sim b,$ i.e. scales as $\sqrt{q}$, which justifies our neglect of $\mathcal{O}(q)$ term in the right hand side of equation \eqref{dPhidt} at the beginning of our analysis.

\begin{figure}
\centering
\includegraphics[width=0.5\textwidth]{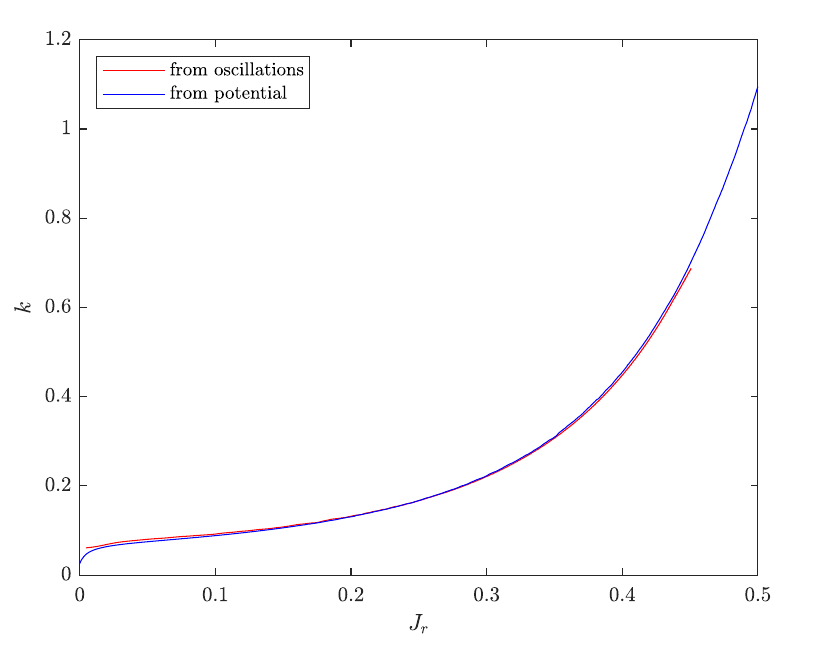}
\caption{A plot showing calculated values of $k(J_{r})$, both from expanding
the potential (blue), and the autoresonant oscillations' frequency, as
derived in equation \eqref{autoresonant frequency}}.
\label{fig:k_coefs}
\end{figure}

For testing this theory we performed full numerical simulations to find the autoresonant oscillations frequency, which on using  equation \eqref{autoresonant frequency}, yields $k(J_{r})$. The latter can be compared with the values obtained from the expansion of the perturbing potential in $\Phi $. This comparison is presented in Fig. \ref{fig:k_coefs} showing a good agreement.

Lastly, note that the eccentricity of the test mass' orbit approaches 1
as $J_{\varphi }\rightarrow 0$ and the test mass approaches the vicinity of the SMBH. In this case, from equation \eqref{cons}, $%
J_{r}\rightarrow 1/p$ and from equation \eqref{Omega_K_AAV}, we have $%
\mbox{${\Omega}_{\rm K}$}\rightarrow p^{3}$. Then, using the resonance
condition \eqref{res_cond1}, we find the limiting driving frequency $%
\mbox{$\omega_{\rm d}$}=(p-1)p^{2}$ for reaching the vicinity of the SMBH.

\section{Astrophysical Implications}
\label{sec:astro}
The theory developed in prior sections will be most applicable to dynamically cold disks of stars surrounding a SMBH in a galactic nucleus.  As mentioned in Sec. \ref{sec:into}, such disks arise naturally in the aftermath of AGN episodes: after the gas is accreted or blown away, young, thin disks of stars on quasi-circular orbits are left behind \citep{Levin&Beloborodov}, and may also naturally host IMBHs \citep{GoodmanTan04}.  

While analytic and semi-analytic models exist to estimate the surface number density $S_\star(R)$ of stars formed by Toomre instability in AGN disks \citep{Thompson, Gilbaum}, these distributions may be sculpted by stellar migration within the disk due to hydrodynamic interactions, which carries significant uncertainties \citep{Grishin+23}.  $S_\star(R)$ may be further modified by capture (due to hydrodynamic drag) of stars from the pre-existing nuclear star cluster, from initially inclined orbits \citep{SyerRees91}.  Despite these substantial uncertainties, we can build an explicit model for $S_\star(R)$ by assuming (i) a Shakura-Sunyaev structure \citep{ShakuraSunyaev73} for the inner regions of the AGN disk; (ii) a steady-state inward flux of migrating stars, $\dot{M}_\star$, which is a fixed fraction $f_\star \ll 1$ of the inward gas accretion rate (i.e. $\dot{M}_\star = f_\star \dot{M}_{\rm g}$); (iii) migration rates that are a combination of ``Type I'' torques $\Gamma_{\rm I}$ \citep{GoldreichTremaine80} and GW torques $\Gamma_{\rm GW}$ \citep{Peters}, producing a total (specific) migratory torque $\Gamma = \Gamma_{\rm I} + \Gamma_{\rm GW}$, where
\begin{equation}
    \Gamma_{\rm I} = C_{\rm I} q_\star^2 \Sigma R^4 \Omega^2 \left(\frac{H}{R}\right)^{-2}. 
\end{equation}
Here we consider the migration of stars with individual masses $m_\star$ and mass ratios $q_\star = m_\star / m_1$ with respect to the central MBH.  The Type I torques produced by the hydrodynamic response of the gas disk to stellar gravity depend on the disk's gas surface mass density $\Sigma$ and aspect ratio $H/R$, hydrodynamic variables for the Shakura-Sunyaev disk model that we define in Appendix \ref{app:disk}.  We assume in this section that the orbital frequency is Keplerian, $\Omega(R) = \sqrt{Gm_1/R^3}$.  The dimensionless prefactor $C_{\rm I}$ can be approximated as \citep{Paardekooper+11}:
\begin{equation}
    C_{\rm I} = 0.8 + 1.0 \xi +0.9 \delta , 
\end{equation}
where $\xi = \partial \ln T / \partial \ln R$ and $\delta = \partial \ln \Sigma / \partial \ln R$ are the local power-law slopes of the gas temperature and surface density profiles, respectively.

Likewise, the specific migratory torque
\begin{equation}
    \Gamma_{\rm GW} = -\frac{32}{5} \frac{G^{7/2}m_1^2 m_\star (m_1+m_\star)^{1/2}}{c^5 R^{7/2}},
\end{equation}
results in a GW inspiral time
\begin{equation}
    t_{\rm GW} = \frac{5 a^4 c^5}{256 G^3 m_1 m_2 m} \label{eq:GW}
\end{equation}
for circular orbits.  We only specify torques (and not dissipation rates) because orbital migration quickly damps out eccentricity, producing quasi-circular inspirals.

Our three underlying assumptions listed above become less accurate as one moves to large radii $R$, where the gas disk structure will be heavily modified by Toomre instability and feedback \citep{Thompson, Gilbaum}, where stellar fluxes are unlikely to be in a steady state \citep{Gilbaum}, and where additional types of migratory torques may become relevant \citep{Grishin+23}.  However, as we shall see, it is the smaller radii that produce the most directly observable signatures of resonance capture, in the form of elevated TDE rates, motivating these assumptions.  

\begin{figure}
\centering
\includegraphics[width=0.5\textwidth]{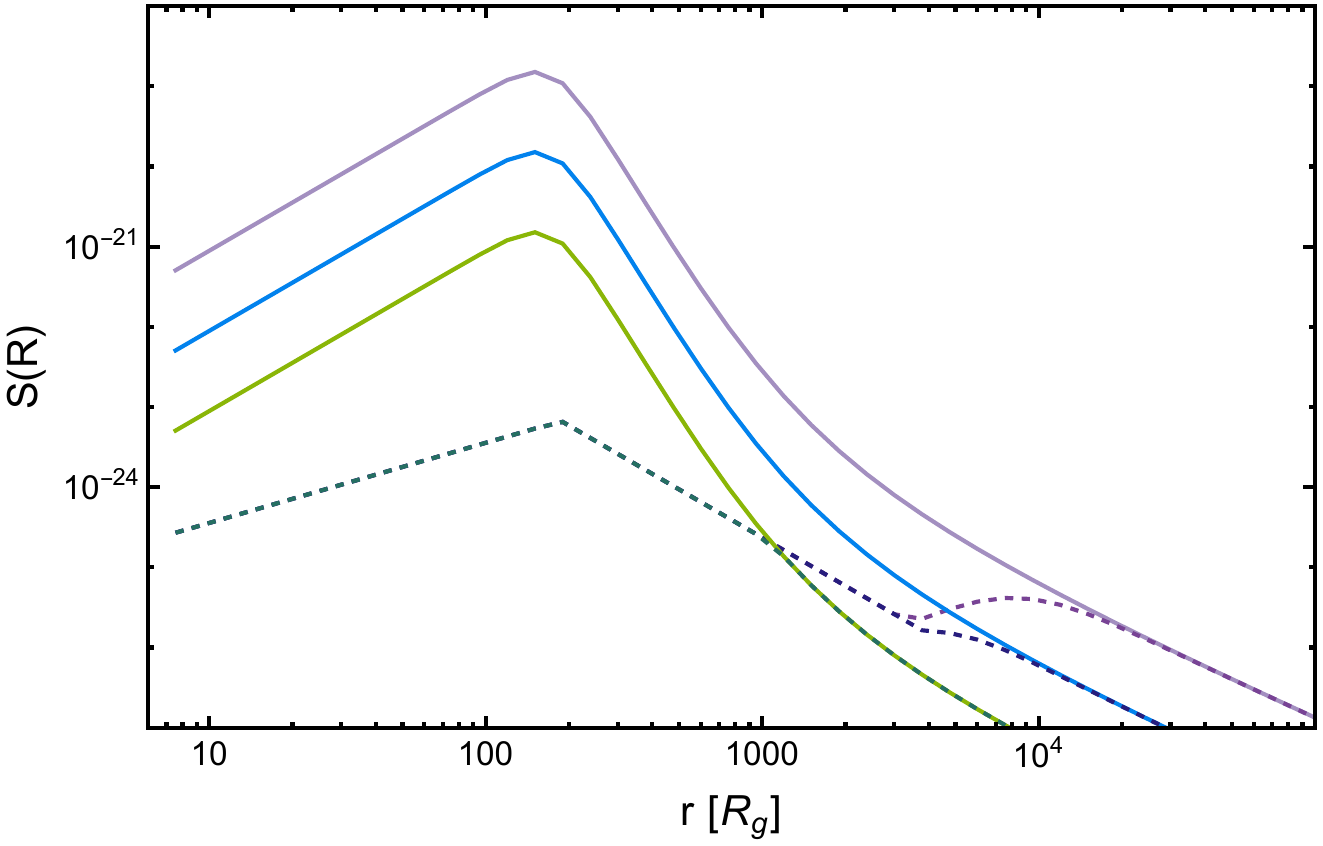}
\caption{The surface density $S(R)$ of stars located at a physical radius $R$, plotted against dimensionless radius $r=R/R_{\rm g}$.  These results assume a past Shakura-Sunyaev gas disk model accreting onto a $M=10^7 M_\odot$ MBH at $\dot{M}=0.1\dot{M}_{\rm Edd}$, and inflow equilibrium for the stellar populations that migrate through the gas disk (at a rate $\dot{M}_\star$) prior to its dissolution.  Purple, blue, and green curves show solutions for $\dot{M}_\star=\{10^{-3}, 10^{-4}, 10^{-5}\}~{\rm yr}^{-1}$, respectively.  After the gas dissipates, the stellar population is eroded through collisions and GW migration as described in the text.  Solid lines show stellar surface densities ignoring these erosion effects, while dashed lines account for them.}
\label{fig:NProfiles}
\end{figure}

In Fig. \ref{fig:NProfiles}, we plot example $S_\star(R)$ distributions as functions of cylindrical radius $R$ in a thin stellar disk.  We show two families of such distributions: first, a naive one, in which a constant inwards flux of stars is computed in the continuum limit and determined by the steady state condition:
\begin{equation}
    S_{\rm ss} = \frac{\dot{M}_\star}{2 \pi m_\star R \dot{a}_\star},
\end{equation}
where $\dot{a}_{\star} = \sqrt{Gm_\star  R}/\Gamma$.  These naive $S(R)$ distributions are strongly peaked at small radii, near the radius where radiation pressure gives way to gas pressure in the progenitor gas disk.

After the gas dissipates, the ``left-behind'' disk of stars will cease its hydrodynamic migration, but $S_\star(R)$ will continue to evolve due to (i) GW inspiral, and (ii) direct physical collisions between stars, which rapidly deplete the disk population on scales $\sim 10^{2-3} R_{\rm g}$.  As star-star collision times are far shorter than GW inspiral times for distant IMBHs, we assume that shortly after the AGN episode ends, stars at semimajor axis $a$ will be radially separated by roughly their own Hill sphere distance, $r_{\rm H} = a_\star (m_1 / m_\star)^{1/3}$.  An alternative possibility is that the stars may assemble into a resonant chain \citep{Secunda+19}, which could have a higher or lower surface density depending on the value of $q_\star$.  Under our assumption of Hill sphere separations, we wind up with a 2D surface number density 
\begin{equation}
    S_{\rm coll}(R) = \frac{1}{2\pi R^2} \left( \frac{m_1}{m_\star} \right)^{1/3}.
\end{equation}
In practice, we take $S_\star(R) = \min(S_{\rm ss}, S_{\rm coll})$ as a realistic stellar disk surface density.

If an IMBH has formed at larger regions in the AGN disk \citep{Goodman&Tan, McKernan+12, Secunda+20}, its inward migration during the AGN episode may be heavily slowed by opening a gap \citep{Kanagawa+18}, allowing it to survive until the gas has dissipated.  Its subsequent GW-driven inspiral will capture populations of disk stars into autoresonance; as we have shown above, the capture probability will be $100\%$ for the $2:1$ resonance and initially circular orbits.  The growth of stellar eccentricity following resonant capture will lead the individual stars to tidally disrupt, at a rate set predictably by Eq. (\ref{eq:GW}).  We plot the resulting TDE rate in Fig. \ref{fig:TDERates}, showing how different $S_{\rm ss}$ models produce very high TDE rates in the decades and centuries prior to merger, but when a more realistic $S_\star(R) \approx S_{\rm coll}(R)$ stellar profile is used, we obtain TDE rates $\sim 10^{-2} - 10^{-1}~{\rm yr}^{-1}$ in the final decade before the SMBH merger.

\begin{figure}
\centering
\includegraphics[width=0.5\textwidth]{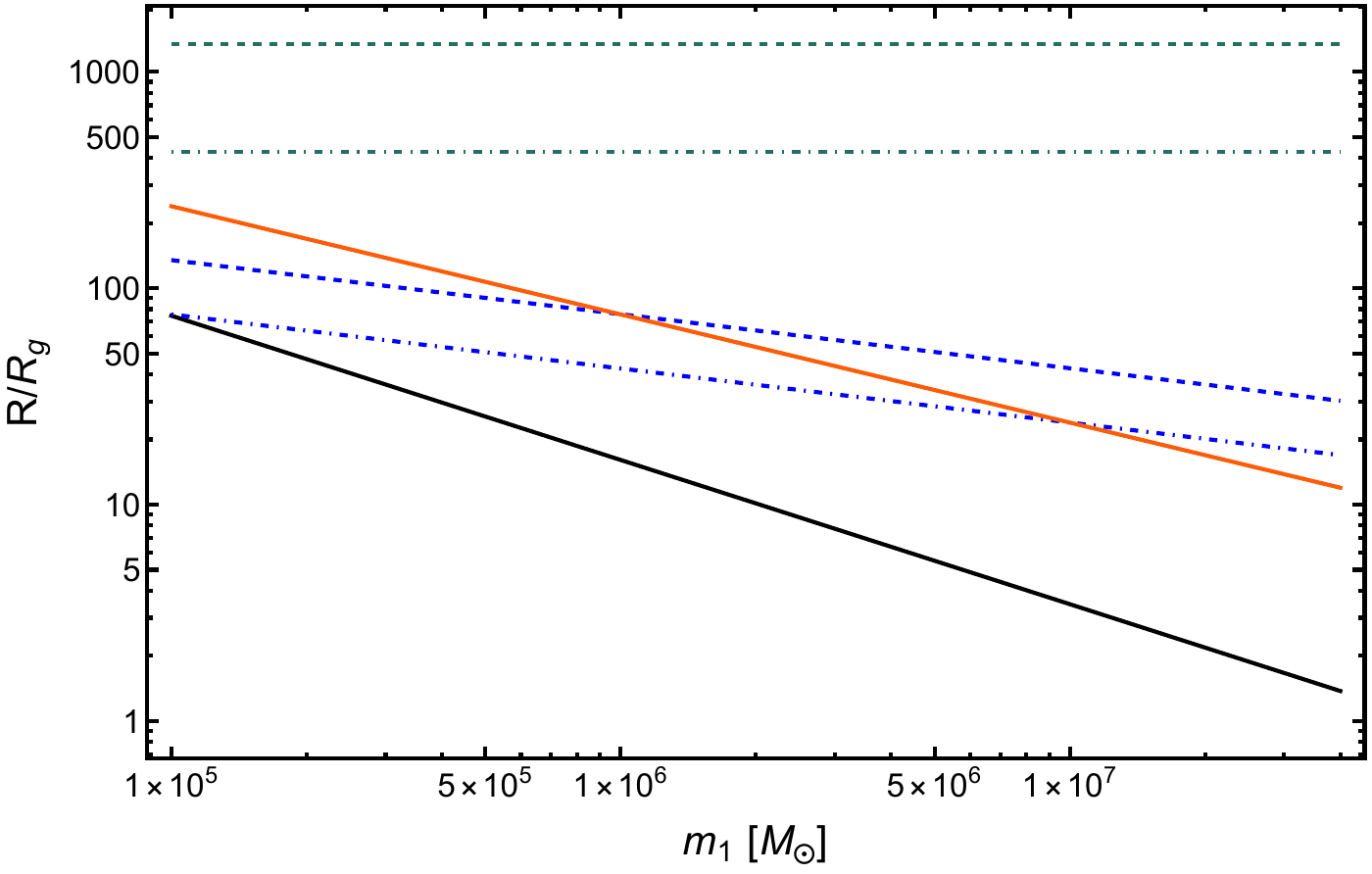}
\caption{Radial scales of interest (normalized to the central gravitational radius $R_{\rm g}$) plotted against primary mass $m_1$.  The tidal disruption radius for a $m_\star = 0.1 M_\odot$ star is shown as a black line.  Blue curves show $a_{10}$ for $q=0.01$ (dot-dashed) and $q=0.001$ (dashed).  The orange curve shows the stellar $a_{\rm GW}$ value for a system age of $t_{\rm age}=10^6 ~{\rm yr}$, and the green curves show the maximum starting radius $a_{\rm IMBH}$ for $q=0.01$ (dot-dashed) and $q=0.001$ (dashed) IMBHs.  }
\label{fig:radii}
\end{figure}

\begin{figure}
\centering
\includegraphics[width=0.5\textwidth]{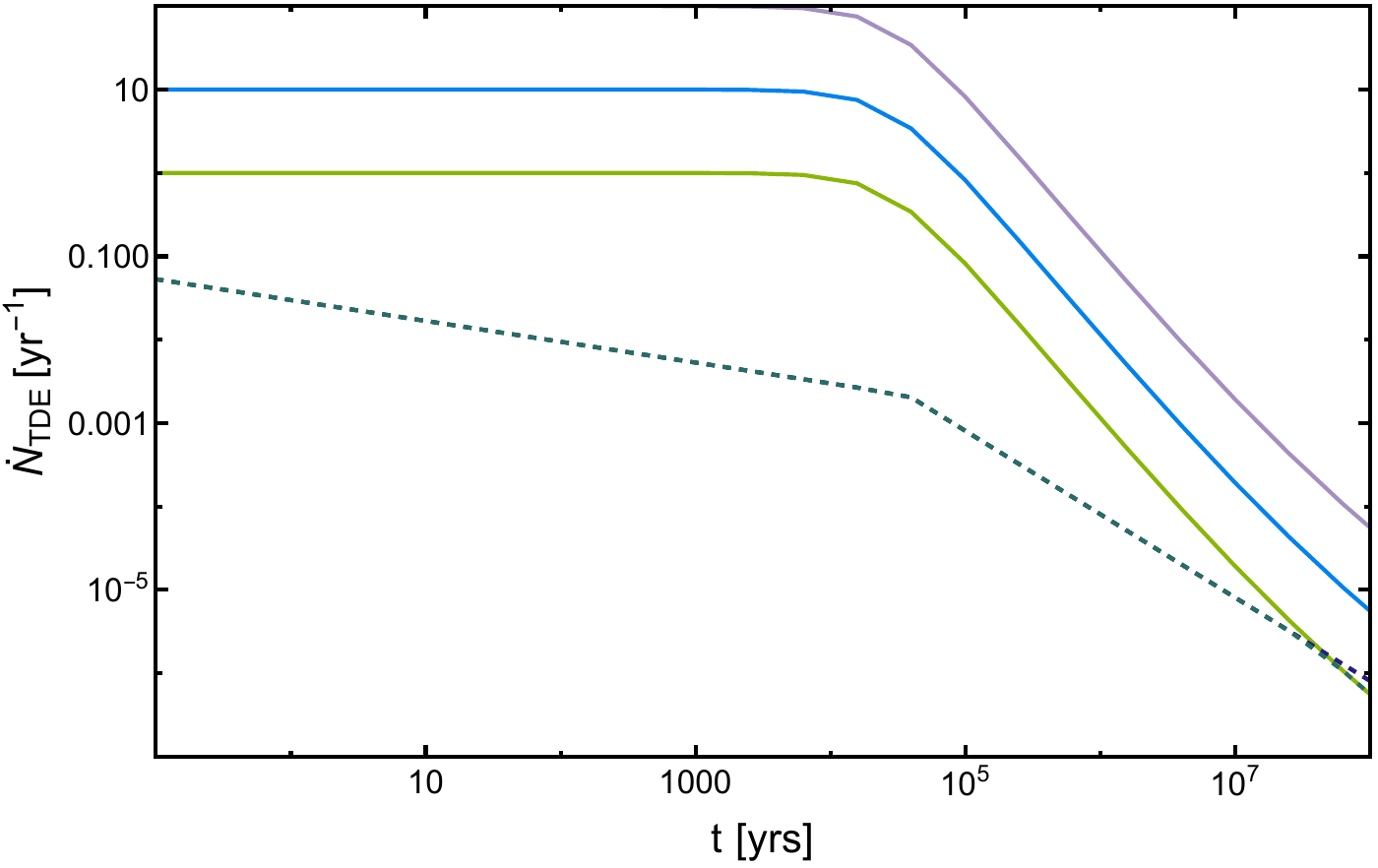}
\caption{TDE rates $\dot{N}_{\rm TDE}$ plotted against the time until merger, $t$, for an SMBH-IMBH binary with $q=10^{-3}$ and the same stellar disk models as in Fig. \ref{fig:NProfiles}.  The dashed line represents the most realistic model for the density profile of the stellar disk (i.e. one where collisions between stars erode the initial disk left behind by star formation in a fragmenting AGN.  Peak TDE rates as high as $\dot{N} \sim 10^{-1} - 10^{-2} ~{\rm yr}^{-1}$ can be achieved for realistic parameters in the final decade before merger.}
\label{fig:TDERates}
\end{figure}

We note that even the eroded stellar densities of $S_\star(R)$ may overestimate the true population of stars at small radii if a long enough period of time has passed, so that the stars themselves undergo a GW-driven inspiral (possibly terminating their lives as some type of quasi-periodic eruption, e.g. \citealt{Metzger+22, LinialSari23}).  For a post-AGN stellar disk of age $t_{\rm age}$, we compute the critical semimajor axis that an object of mass $\mathcal{M}$ needs to survive:
\begin{equation}
a_{\rm GW} = \left( \frac{256 G^3 m_1 \mathcal{M} (m_1 + \mathcal{M}) t_{\rm age}}{5 c^5} \right)^{1/4}.
\end{equation}
By setting $\mathcal{M} = m_2$ and $t_{\rm age} = 10~{\rm yr}$, we compute $a_{10}$, the semimajor axis of the IMBH one decade prior to merger.  We then find the system age that gives $a_{\rm GW}(\mathcal{M}=m_\star) = a_{10}$; this critical system age $t_{\rm crit} = 10~{\rm yr}~ m_2 / m_\star$.  Finally, we use this system age to compute $a_{\rm IMBH}$, the maximum initial semimajor axis of the IMBH at the end of the AGN episode; IMBHs that begin at larger radii will not have enough time to migrate into the 2:1 resonance before the left-behind stars are depleted by GW inspiral.  The simple result of this is that 
\begin{equation}
    a_{\rm IMBH} / R_{\rm g} \approx 1350 \left( \frac{q}{0.01} \right)^{1/2} \left( \frac{m_\star}{0.1 M_\odot} \right)^{-1/4}.
\end{equation}
This is a sufficiently large radius to be a plausible site for either IMBH formation (in e.g. a migration trap) or for the IMBH to have stalled hydrodynamically due to gap opening \citep{Grishin+23}.  The relevant radial scales are shown in Fig. \ref{fig:radii} for a $m_\star = 0.1 M_\odot$ red dwarf.  We see that there is little parameter space for autoresonant capture when $m_1 \ll 10^6 M_\odot$, but the process is feasible for supermassive MBH primaries.

Although the model presented above gives us a simple way to estimate surface densities of disk stars left behind after a major AGN episode, and the TDE rates that result from resonant capture by an inward migrating IMBH, it contains a number of significant assumptions, which we summarize here:
\begin{itemize}
    \item Migration of stars (and compact objects) through the gas disk is due purely to Type I hydrodynamic torques and GW dissipation.  This model neglects the Type II \citep{Kanagawa+18} migration regime (which is unlikely to be triggered in much of the AGN parameter space; \citealt{Gilbaum}) and the substantially more uncertain thermal torques.  Importantly, we assume that no migration traps (i.e. regions where $\Gamma_{\rm tot} > 0$ exist in the disk, as these would make a steady-state stellar inflow solution of our type impossible to achieve.  The existence of migration traps in AGN disks has been predicted in the past \citep{Bellovary+16}, but the most recent calibrations of Type I torques \citep{JimenezMasset17} are incompatible with such traps on the small scales relevant for us \citep{Grishin+23}.
    \item Stars move through the disk on quasi-circular, prograde orbits.  This is likely a good assumption for stars that form {\it in situ} \citep{Stone1}, or those that capture into the disk on prograde orbits (and circularize/align).  However, we neglect stars that capture on retrograde orbits, as (i) these will not possess orbital resonances with the outer MBH binary, and (ii) they may be quickly destroyed due to eccentricity excitation from gas dynamical friction \citep{Secunda+21}.
    \item A steady state solution is established, where the flux of stars through the inner disk, $\dot{M}_\star$, equals some quasi-steady rate of star formation and capture in the outer disk.  Such a steady state could be complicated by vigorous accretion feedback from embedded compact objects \citep{Chen+23}, but that is beyond the scope of this work.
    \item The gas dissipates fairly quickly throughout the entire AGN disk, so that the steady-state inflow solution $S_\star(R)$ ``freezes in'' as the AGN episode ends.  To the extent that this assumption is violated, it will cause deviations from our $S_\star(R)$ profiles in regions with the shortest migration times $t_{\rm mig}$; these are generally regions with smaller $S_\star(R)$ densities and (as we shall see) lower contributions to the TDE rate.
\end{itemize}

In summary, the final few decades of SMBH-IMBH inspiral through the disk of stars left behind after an AGN episode will see TDE rates rise several orders of magnitude above their values in normal galactic nuclei (where $\dot{N}_{\rm TDE} \sim 10^{-5} - 10^{-4}~{\rm yr}^{-1}$; \citealt{Stone2, Yao+23}).  These rates are high enough that a significant percentage of {\it LISA}-band mergers (tens of percent) from such systems will see an individual TDE flare in the final decade before {\it LISA} detection.  This type of electromagnetic counterpart opens the door to host galaxy identification, the implications of which we will discuss in the final section.

\section{Conclusions}
\label{sec:conclusions}
We have studied the resonant dynamics of stars in the vicinity of a hierarchical massive black hole binary and discussed the resulting TDE rate into the central SMBH in the binary. Our fiducial calculations were motivated by disks of stars that form from Toomre instability in AGN episodes \citep{Levin&Beloborodov, Levin07, Gilbaum}, as well as IMBHs that may also arise \citep{Goodman&Tan, McKernan+11} in these disks.  The simple geometry of this scenario allowed us to use the planar restricted three-body limit to analyze the capture of stars into resonance 
as the MBH binary decays from gravitational wave losses.  For stars that (i) are well interior to the binary orbit and (ii) are initially on nearly circular orbits, capture into the 2:1 resonance is guaranteed if the chirp rate of the Keplerian frequency of the secondary BH is below a threshold that is always satisfied for astrophysical inspirals. After the capture into resonance, the stars stay in resonance continuously and increase their orbital eccentricity towards unity \citep{YuTremaine01}.  The drop in the test particle's pericenter distance eventually leads to a TDE, provoking a bright electromagnetic outburst accompanying the MBH binary inspiral \citep{Seto&Muto1,Seto&Muto2}.  This {\it precursor} TDE can serve as an electromagnetic counterpart distinct from {\it postcursor} TDEs which may be expected due to GW recoil after the MBH merger \citep{StoneLoeb11, AkibaMadigan21}

This autoresonant process is similar to that in the formation of Plutinos in the Kuiper Belt in early evolution of the Solar system \citep{Friedland}. However, in contrast to the Plutinos, which capture into outer mean motion resonances, the capture of stars into autoresonance by MBH binaries involves inner resonances \citep{YuTremaine01}. The TDEs produced by autoresonance have a very different dynamical origin than the classical ``loss cone'' picture, in which weak, diffusive star-star scatterings cause a slow random walk in stellar orbits up to high eccentricity. Since this novel channel of TDEs occurs on the time scale of the MBH inspiral and involves a large number of stars in autoresonance, we predict much higher rates of disruptions than the classical channel. The peak rates we find in the final decade of an SMBH-IMBH inspiral can be as high as $\dot{N}_{\rm TDE} \sim 10^{-1}~{\rm yr}^{-1}$, far greater than the $\dot{N}_{\rm TDE} \sim 10^{-4}~{\rm yr}^{-1}$ predicted by loss cone theory \citep{Stone2} or the even lower $\dot{N}_{\rm TDE} \sim 10^{-5}~{\rm yr}^{-1}$ found by current optical surveys of TDEs.  The elevated rate suggests that some galactic nuclei may exhibit multiple TDEs, creating a possible tracer of the binary MBH population.  This channel will also give rise to multi-messenger observations: an electromagnetic flare from the TDEs that accompanies or precedes the {\it mHz} gravitational wave signal from an inspiraling SMBH-IMBH binary.  If such a multimessenger signal could be detected, it would be of extreme astrophysical value: {\it LISA} sky error regions are too large to permit host galaxy localization, but if the host can be identified, the luminosity distance measurement from GWs could be combined with an electromagnetic redshift to enable ``standard siren'' cosmological measurements \citep{Schutz86}.

In our analysis we have made several simplifying assumptions. The first and most obvious one is the planarity of the problem -- we have studied a system in which both the MBHs, and the test stars, all orbit in a single plane. While this assumption is well-motivated for the ``left-behind'' stellar disks following AGN episodes, many other MBH binaries form in quasi-spherical nuclear star clusters \citep{Milosavljevic&Merritt01, Merritt&Milosavljevic05}.  In a quasi-spherical system, generic stellar orbits will initially be both misaligned and eccentric, a more complicated problem that can be explored by numerical integration \citep{NamouniMorais15}.  A second assumption concerning system parameters is that of an extreme mass ratio ($q \ll 1$) between the primary SMBH and the secondary IMBH.  While, again, this is reasonably well motivated for IMBHs that form in AGN disks, such extreme mass ratios will be uncommon for MBH binaries produced by galaxy mergers, as the dynamical friction time for a very low-mass satellite can exceed the age of the Universe \citep{Taffoni+03}.  As $q$ increases, non-linear effects may destabilize autoresonant capture, and it is likely that capture probabilities will need to be assessed by numerical simulations.

Other assumptions we have made concern the underlying physics rather than the parameters of the problem.  For example, we treat gravity as being purely Newtonian for the test particle (though of course it is relativistic radiation reaction that powers the MBH inspiral). The leading order impact of general relativity is to induce apsidal precession, which may affect the stability of resonances as eccentricity increases.  We have also considered single star dynamics only, disregarding star-star scattering events.  Stellar scattering may reduce TDE rates by knocking stars out of autoresonance; on the other hand, if relativistic apsidal precession causes a high-eccentricity exit from resonance, a very small amount of star-star scattering may cause these stars to random walk into a TDE.  Finally, we have neglected the population of higher order resonances (e.g. 3:1 or 4:1) that stars will encounter before the 2:1 resonance.  However, capture into these resonances at mass ratios $q\ll 1$ studied here is probabilistic rather than deterministic, so we expect that an $\mathcal{O}(1)$ fraction of disk stars will pass through higher order resonances without a major impact on their orbit before encountering the 2:1 resonance.

In summary, autoresonant capture of stars by inspiraling MBH binaries is capable of producing bursts of TDEs during the final stages of MBH mergers.  While this first investigation shows that these TDE bursts have potential as electromagnetic counterparts to {\it LISA}-band signals, their overall relevance must be quantified by further analytic and numerical studies relaxing some of the assumptions listed above.

\bibliographystyle{mnras}
\bibliography{refs}

\appendix
\onecolumn
\section{Action-Angle Variables}
\label{app:AAV}
Here, for completeness, we summarize the developments yielding the transformation to AA variables in our problem. The planar elliptical Keplerian motion conserves the energy $H_{0}$ and is periodic in both $%
\varphi $ and $r$, such that in $\varphi $ it conserves the angular momentum 
$p_{\varphi }$, while the motion in $r$ can be viewed as an oscillation in
the one-dimensional potential well $V=p_{\varphi }^{2}/(2r^{2})-1/r$. This
suggests to define the following azimuthal and radial action variables \citep{Landau&Lifshitz}.

\begin{equation}
J_{\varphi }=\frac{1}{2\pi }\int_{0}^{2\pi }p_{\varphi }d\varphi =p_{\varphi
}  \label{A11}
\end{equation}%
and%
\begin{equation}
J_{r}=\frac{2}{2\pi }\int_{r_{\min }}^{r_{\max }}\sqrt{2(H_{0}-V_{eff})}%
dr=-J_{\varphi }+(2\left\vert H_{0}\right\vert )^{-1/2}.  \label{A2}
\end{equation}%
The last \ equation yields relation of the Hamiltonian to the actions%
\begin{equation}
H_{0}(J_{r},J_{\varphi })=-\frac{1}{2(J_{r}+J_{\varphi })^{2}}.  \label{A3}
\end{equation}%
It can be also shown that in the elliptical orbit equation
\begin{equation}
\frac{\ell}{r}=1-e\cos{\varphi}.  \label{orbit}
\end{equation}
parameter $\ell$ and
eccentricity $e$ are simply related to the two actions \citep{Landau&Lifshitz}:

\begin{equation}
\ell=J_{\varphi }^{2},e^{2}=1-\frac{J_{\varphi }^{2}}{(J_{r}+J_{\varphi })^{2}},
\label{A4}
\end{equation}%
which also yields relations of the major and minor semi axis $%
a=\ell/(1-e^{2})=(J_{r}+J_{\varphi })^{2}$ and $b=\ell/\sqrt{1-e^{2}}=J_{\varphi
}(J_{r}+J_{\varphi })$ to the actions. Finally, we complete the canonical
transformation $(p_{r},p_{\varphi },r,\varphi )\rightarrow (J_{r},J_{\varphi
},\Theta _{r},\Theta _{\varphi })$ to AA variables by defining the
generating function 
\begin{equation}
W(r,\varphi ,J_{r},J_{\varphi })=\varphi J_{\varphi }+\int^{r}Pdr'  \label{A5}
\end{equation}%
where $P=\pm \sqrt{2[H_{0}(J_{r},J_{\varphi })-J_{\varphi
}^{2}/(2r^{2})+1/r]}$. This function gives the required four transformation
equations 
\begin{eqnarray}
\frac{\partial W}{\partial r} &=&P=p_{r}  \nonumber \\
\frac{\partial W}{\partial \varphi } &=&J_{\varphi }=p_{\varphi }  \nonumber
\\
\frac{\partial W}{\partial J_{r}} &=&\int^{r}\frac{\partial H_{0}}{\partial
J_{r}}\frac{dr'}{P}=\Theta _{r}  \label{A6} \\
\frac{\partial W}{\partial J_{\varphi }} &=&\varphi -R=\Theta _{\varphi } 
\nonumber
\end{eqnarray}%
where
\begin{equation}
R=\int^{r}\left( \frac{J_{\varphi }}{r'^{2}}-\frac{\partial H_{0}}{\partial
J_{\varphi }}\right) \frac{dr'}{P}
 \label{R}%
\end{equation}
The first two equations in this set are natural connections to the radial
and azimuthal momenta, while the last two equations define the canonical
angles $\Theta _{r,\varphi }$. Finally, note that since the Hamiltonian $H_{0}
$ does not depend on the angle variables both actions remain constant, while Hamilton equations for the canonical angles are 
\begin{equation}
\frac{d\Theta _{r}}{dt}=\frac{d\Theta _{\varphi }}{dt}=\Omega _{K}
\label{A7}
\end{equation}%
where $\OmgK$ -- the Keplerian frequency -- is defined as
\begin{equation}
\Omega _{K}=\frac{1}{(J_{r}+J_{\varphi })^{3}}. \label{OmK}
\end{equation}
We complete this appendix by discussing the limit of small oscillations $%
\delta r$ around the initially circular orbit of unit radius, i.e., $r=1+\delta$, with angular velocity equal to unity. These are \textit{linear} oscillations in the
potential well $V$ mentioned above, and for small action $J_{r}$ can be expressed as $\delta=\sqrt{2J_{r}}\cos \Theta _{r}$, where the amplitude $A$ of the oscillations, as for all linear oscillators, is related to the action via $J_{r}=\Omega
_{K}A^{2}/2\approx A^{2}/2$. In addition, for small oscillations, since $P=dr/dt$ is the velocity in the potential well,
\begin{equation}
R=\int^{t}\left( \frac{J_{\varphi }}{r^{2}}-\Omega _{K}\right) \dt\approx
\int^{t}[J_{\varphi }(1-2\delta)-\Omega _{K}]\dt\approx -2\int^{t}\delta
\dt=2\sqrt{2J_{r}}\sin \Theta _{r}  \label{A8}
\end{equation}%
and thus%
\begin{equation}
\varphi =\Theta _{\varphi }+2\sqrt{2J_{r}}\sin \Theta _{r} . \label{A9}
\end{equation}%
This result will be used in Appendix B.

\section{Effective potential for Small Eccentricities}
\label{app:small_AR}
In this appendix, we will focus on the case of small eccentricities. In this case, using the single resonance approximation \citep{Chirikov}, we leave a single $p$-th harmonic term in the perturbing potential 
\eqref{per_poten_harm}, i.e. 
\beq{}
    V_p = \frac{1}{r_2}f_p(\alpha)\cos{(p\theta)}.
\eeq
Here we write (see the end of \ref{app:AAV}) $\alpha=r_2/r\approx r_2/(1+\delta)$, where $\delta = \sqrt{2J_r}\cos{\Thr}\ll1$. Then, by expanding to first order in $\delta$, we get
\beq{}
     V_p \approx \frac{1}{r_2}f_p(r_2-\delta r_2)\cos{(p\theta)} \approx\left(\frac{1}{r_2}f_p(r_2)-\delta \frac{\p f_p}{\p \alpha}\right)\cos{(p\theta)},
\eeq
where initially, $r=1$ and $\alpha=r_2$. Next, we substitute equation \eqref{A9} into $V_p$ and to get
\beq{}
    V_p = \left(M_p-\delta N_p \right) \cos{\left[p(\Thphi-R-\psi)\right]},
\eeq
where $M_p = f_p(r_2)/r_2$ and $N_p = \p f_p/\p\alpha$. Here we use equation \eqref{A8} for $R$, to obtain
\beq
    V_p=M_p\cos{\left(p(\Theta_{\varphi}-\psi)\right)}+ \sqrt{2J_r}\left(M_p\cos{\Thr}\cos{\left(p(\Thphi-\psi\right)}+2pN_p\sin{\Thr}\sin{\left(p(\Thphi-\psi)\right)}\right) + \mathcal{O}(J_r).
\eeq
Alternatively,
\beq
    V_p=M_p\cos{\left(p(\Theta_{\varphi}-\psi)\right)} + \sqrt{2J_r}\left(\frac{M_p}{2}-pN_p\right)\cos{\left(p(\Thphi-\psi)+\Thr\right)} + \sqrt{2J_r}\left(\frac{M_p}{2}+pN_p\right)\cos{\left(p(\Thphi-\psi)-\Thr\right)}.
\eeq
Finally, we leave the only resonant (last) term in this equation, to obtain
\beq
    V_p\approx -\sqrt{2J_r}\zeta_p\cos{\Phi},
\eeq
where $\zeta_p = M_p/2+pN_p$ and $\Phi=p(\Thphi-\psi)-\Thr-\pi$. 
The first three coefficients $\zeta_p$ in $V_p$ are $\zeta_2=0.750$, $\zeta_3=1.546$, and $\zeta_4=2.345$.

\section{Derivation of the resonant perturbing Potential}
\label{app:potential_derivation}

Here, we derive the resonant perturbing potential in our problem
for arbitrary radial action $J_{r}$. We proceed from equation %
\eqref{per_poten_harm} and substitute $\varphi =\Theta _{\varphi }-R$ \eqref{A6}:%
\begin{equation}
{}V_{1}=\frac{1}{r_{2}}\sum_{j=0}^{\infty }f_{j}\cos {\left( j(\Thphi-R)-\psi )\right) }.  \label{poten_AAV}
\end{equation}%
This can be rewritten as 
\begin{equation}
{}V_{1}=\sum_{j=0}^{\infty }A_{j}\cos {\left( j(\Thphi-\psi)\right) }+\sum_{j=0}^{\infty }B_{j}\sin {\left( j(\Thphi-\psi )\right) },  \label{poten_AAV_sep}
\end{equation}%
where $A_{j}=f_{j}\cos {(jR)}/r_{2}$ and $B_{j}=f_{j}\sin {(jR)}/r_{2}$.
The goal is to extract the resonant part in $V_{1}$ for arbitrary $J_{r}$, similar to the development in \ref{app:small_AR} for small $J_{r}$. We'll again focus on $p-1:p$ resonance ($p\in \mathbb{N}$), i.e. assume that the system continuously preserves 
\begin{equation}
{}(p-1)\OmgK-p\omd\approx 0,
\label{res_cond}
\end{equation}%
or equivalently, the phase mismatch $\Phi =p\Thphi-\Thr +p\psi-\pi $
remains nearly stationary for all $J_{r}$ during the autoresonant evolution. Then, we expand $A_{j}$
and $B_{j}$ in Fourier series 
\begin{equation}
A_{j}=\sum_{m=1}^{\infty }a_{j}^{m}(J_{r})\cos {(m\Thr)}%
;B_{j}=\sum_{m=1}^{\infty }b_{j}^{m}(J_{r})\sin {(m\Thr)},
\label{a_b}
\end{equation}%
allowing to write the perturbing potential as 
\begin{equation}
{}V_{1}=\sum_{j=0}^{\infty }\sum_{m=1}^{\infty }V_{j}^{m},  \label{V1}
\end{equation}%
where 
\begin{equation}
{}V_{j}^{m}=a_{j}^{m}\cos {(m\Thr)}\cos {[j(\Thphi-\psi )] }+b_{j}^{m}\sin {(m\Thr)}%
\sin {[j(\Thphi-\psi )] }.  \label{Vjm_expanded}
\end{equation}%
Alternatively, 
\begin{equation}
{}V_{j}^{m}=c_{j}^{m}\cos {[j(\Thphi-\psi )-m\Thr-\pi]}+d_{j}^{m}\cos {[j(\Thphi-\psi )+m\Thr-\pi]},  \label{Vjm}
\end{equation}%
where $c_{j}^{m}=-(a_{j}^{m}+b_{j}^{m})/2$ and $%
d_{j}^{m}=-(a_{j}^{m}-b_{j}^{m})/2$. In seeking the resonant contribution, we
choose $j=mp$ in equation \eqref{Vjm} because the coefficients fall rapidly  and leave the first cosine term only to get the resonant perturbing potential 
\begin{equation}
{}V_{p}(J_{r},\Phi )\approx \sum_{m=1}^{\infty }c_{mp}^{m}(J_{r})\cos {%
\left( m\Phi \right) }.  \label{V_p}
\end{equation}%
The coefficients $c_{mp}^{m}$ can be found by using the Fourier series: 
\begin{eqnarray}
{}a_{mp}^{m} &=&\frac{1}{2\pi }\int_{0}^{2\pi }f_{mp}\cos {(mpR)}\cos {(m%
\Thr)}\mbox{${\rm d}$}\Thr \\
b_{mp}^{m} &=&\frac{1}{2\pi }\int_{0}^{2\pi }f_{mp}\sin {(mpR)}\sin {(m%
\Thr)}\mbox{${\rm d}$}\mbox{$\Theta_r$},
\end{eqnarray}%
yielding 
\begin{equation}
{}c_{mp}^{m}(J_{r})=-\frac{a_{j}^{m}+b_{j}^{m}}{2}=-\frac{1}{2\pi }\int_{0}^{2\pi }f_{mp}\cos {\left[ m(pR-\Thr)\right] }\mbox{${\rm d}$}\Thr.
\end{equation}%
We calculate these coefficients by defining the function  
\begin{equation}
    C_{mp}^{m}(\Thr)=-\frac{1}{2\pi }\int_{0}^{\Thr}f_{mp}\cos {\left[ m(pR-\Theta_{r}^{\prime})\right] }\mbox{${\rm d}$}\Theta_{r}^{\prime}.
\label{Cmpm}
\end{equation}
By using Eqs. \eqref{Cmpm} and \eqref{R}, to write a system of ODEs 
\begin{eqnarray}
{}\frac{\mbox{${\rm d}$}C_{mp}^{m}}{\mbox{${\rm d}$}\Thr} &=& -\frac{1}{2\pi }f_{mp}(r)\cos {[m(pR-\Thr)]} \\
\frac{\mbox{${\rm d}$}R}{\mbox{${\rm d}$}\Thr} &=&1-\frac{J_{\varphi}}{\OmgK r^{2}},
\end{eqnarray}%
where $\OmgK$ is given by \eqref{OmK}, $J_{\varphi }=1-pJ_{r }$, by the conservation law derived at equation \eqref{cons}, and $r$ is given by the orbit
\eqref{A4} with $\Theta_{\varphi}$ replaced by $\Theta_{r}$
\begin{equation}
r=\frac{J_{\varphi }^{2}}{1-e\cos {(\Thr-R)}}.
\end{equation}%
Numerical solution of this system with $\Theta _{r\text{ }}$from $0$ to $%
2\pi $ yields coefficients $c_{mp}^{m}=C_{mp}^{m}(\Theta _{r\text{ }}=2\pi )$
in the resonant perturbing potential $V_{p}$. In the case of small eccentricities (see appendix B), we found that coefficients $c_{mp}^{m}$ decrease rapidly with $m$ allowing to use only single component with $m=1$ in \eqref{V_p}. For larger eccentricities, $c_{mp}^{m}$ decrease slower and we include additional $m>2$ terms in \eqref{V_p} for convergence.

\section{Astrophysical Disk Model}
\label{app:disk}
The stellar disks whose component stars are the subject of this paper are ultimately produced as leftovers of AGN episodes, where an axisymmetric  disk of gas and plasma orbits a central MBH.  Although the precise details of the gas disk turn out not to matter too much for the distribution of stars far after the end of the AGN episode (see \S \ref{sec:astro}), we include the relevant equations for the structure of the gas disk here, as they are needed to reach this conclusion.

For simplicity, we treat the AGN disk structure in the standard Shakura-Sunyaev way: as an axisymmetric, geometrically thin disk orbiting in the nearly Newtonian potential of a MBH with mass $M$, with hydrodynamics in a steady state inflow-equilibrium \citep{ShakuraSunyaev73}.  The disk is assumed to be sub-Eddington, with no outflows, and the microphysics (opacity, equation of state) are treated in an approximate manner.  The disk profile is defined by a 1D surface mass density $\Sigma(R)$.  At each cylindrical radius $R$ there is a disk scale height $H \ll R$, a midplane temperature $T$, a pressure $P$, a photon optical depth $\tau$, a midplane opacity $\kappa$, a midplane gas sound speed $c_{\rm s}$, and a midplane 3D mass density $\rho$.  Although such a 1D treatment of hydrodynamics is formally laminar, there is also an effective kinematic viscosity $\nu$ which parametrizes the transport of angular momentum by small-scale turbulent motions.  The system of algebraic equations governing the disk structure is
\begin{align}
    \rho =& \Sigma / H \\
    H =& c_{\rm s}/\Omega \\ 
    c_{\rm s}^2 =& \gamma \frac{P}{\rho} \\
    P =& a T^4 + \frac{\rho k T }{\mu m_{\rm p}} \label{eq:Pressure} \\
    \sigma \frac{T^4} {\tau} =& \frac{3 G M \dot{M} f(R)}{8\pi R^3} \\
    \tau =& \Sigma \kappa \\
    \kappa =& \kappa_{\rm es} + \kappa_0 \rho T^{-7/2}\\
    \nu \Sigma=& \frac{\dot{M} f(R)}{3\pi} \\
    \nu =& \alpha c_{\rm s} H
\end{align}
In these equations, we assume nearly Keplerian gas motion at a frequency $\Omega = \sqrt{G M / R^3}$.  The total opacity $\kappa$ is the sum of electron scattering opacity $\kappa_{\rm es}\approx 0.20(1+X)~{\rm cm}^2~{\rm g}^{-1}$ and the Kramers parametrization of bound-free opacity, $\kappa_{\rm K} \approx 3 \times 10^{25}~{\rm cm}^2~{\rm g}^{-1} Z(1+X+3Y/4) (\rho / {\rm g~ cm^{-3}}) (T / {\rm K})^{-7/2}$, where we assume that the hydrogen, helium, and metal mass fractions have Solar-like values of $X = 0.75$, $Y=0.24$, and $Z=0.01$, respectively.  We note that $a$ is the radiation constant, $\sigma$ the Stefan-Boltzmann constant, $k$ the Boltzmann constant, $\gamma$ the gas adiabatic index (which varies between $\gamma=4/3$ and $\gamma=5/3$ according to Eq. \ref{eq:Pressure}), $m_{\rm p}$ is the proton mass, and $\mu \approx 0.6$ is the mean molecular weight.  The dimensionless function $f(R)=1-\sqrt{R_{\rm in}/R} \approx 1$ arises from assuming zero hydrodynamical stress at the inner boundary $R_{\rm in}=6 R_{\rm g}$, and the dimensionless Shakura-Sunyaev viscosity parameter $\alpha$ is set equal to 0.1.  These 9 equations are combined numerically into a single equation which is solved numerically by root-finding to yield disk profiles such as $\Sigma(R)$, $T(R)$, etc.
\end{document}